\newif\ifC
\Cfalse  
\ifC
    \documentclass[a4paper,UKenglish,cleveref, autoref, thm-restate]{lipics-v2021}
\else
    \documentclass{article}
\fi

\ifC
\title{Reading between the levels: new tools for the Thorup--Zwick hierarchy and their applications}
\else
\title{Reading between the levels: new tools for the Thorup--Zwick hierarchy and their applications}
\fi

\ifC
\else
\fi

\ifC
\author{Shiri Chechik}
{Department of Computer Science, Tel Aviv University, Israel}
{shiri.chechik@gmail.com}
{}
{}

\author{Gur Lifshitz}
{Department of Computer Science, Tel Aviv University, Israel}
{gur.lifshitz@gmail.com}
{}
{}

\author{Avi Kadria}
{Department of Computer Science, Bar-Ilan University, Ramat Gan 5290002, Israel}
{avi.kadria3@gmail.com}
{https://orcid.org/0000-0001-8449-3284}
{}

\author{Liam Roditty}
{Department of Computer Science, Bar-Ilan University, Ramat Gan 5290002, Israel}
{liam.roditty@biu.ac.il}
{https://orcid.org/0000-0002-5289-198X}
{Supported in part by BSF grants 2016365 and 2020356.}



\relatedversion{} 

\EventEditors{John Q. Open and Joan R. Access}
\EventNoEds{2}
\EventLongTitle{42nd Conference on Very Important Topics (CVIT 2016)}
\EventShortTitle{CVIT 2016}
\EventAcronym{CVIT}
\EventYear{2016}
\EventDate{December 24--27, 2016}
\EventLocation{Little Whinging, United Kingdom}
\EventLogo{}
\SeriesVolume{42}
\ArticleNo{23}

\else
\author{Shiri Chechik\thanks{Department of Computer Science, Tel Aviv University, Israel. E-mail {\tt shiri.chechik@gmail.com}.} \and
Gur Lifshitz\thanks{Department of Computer Science, Tel Aviv University, Israel. E-mail {\tt gur.lifshitz@gmail.com}.} \and
Avi Kadria\thanks{Department of Computer Science, Bar-Ilan University, Ramat Gan 5290002, Israel. E-mail {\tt avi.kadria3@gmail.com}.} \and Liam Roditty\thanks{Department of Computer Science, Bar-Ilan University, Ramat Gan 5290002, Israel. E-mail {\tt liam.roditty@biu.ac.il}. Supported in part by BSF grants 2016365 and 2020356.}}

\fi

\ifC
\else
\usepackage[margin=1in]{geometry}
\fi
\usepackage{setspace}
\usepackage{multirow}
\usepackage{booktabs}
\usepackage{makecell}
\usepackage[bottom]{footmisc}

\usepackage{epigraph}
\usepackage{graphicx}
\usepackage{tikz}
\usetikzlibrary{arrows,automata,patterns}
\ifC\else\usepackage{enumitem}\fi
\usepackage{hyperref}
\usepackage{lipsum}
\usepackage{amssymb}

\usepackage[linesnumbered,ruled,vlined,boxed,algo2e]{algorithm2e}
\usepackage{nicefrac}
\usepackage{amsmath}
\usepackage{float}
\usepackage{mathtools}
\usepackage{amsthm}
\ifC\else
\usepackage{fullpage}
\fi
\usepackage{amsfonts}
\usepackage{clipboard}
\ifC
\else
\usepackage[nameinlink]{cleveref}
\newtheorem{theorem}{Theorem}

\newtheorem{lemma}{Lemma}

\newtheorem{claim}{Claim}[lemma]

\newtheorem{corollary}[theorem]{Corollary}

\newtheorem{definition}[theorem]{Definition}

\fi
\newtheorem{cclaim}{Claim}[lemma]

\crefname{cclaim}{Claim}{Claim}

\newtheorem{question}{Question}

\Crefname{subclaim}{Claim}{Claims}
\Crefname{claim}{Claim}{Claims}
\crefname{claim}{Claim}{Claims}
\Crefname{question}{Question}{Questions}
\crefname{question}{Question}{Questions}
\Crefname{enumi}{(item)}{(items)}

\newcommand{\Reminder}[1]{

\vspace{0.5em}
\noindent\textbf{Reminder of~\Cref{#1}.} \textit{\Paste{#1}}
\vspace{0.5em}

}

\usepackage{xcolor}
\definecolor{DarkGreen}{RGB}{1,50,32}
\usepackage{pgfplots}
\pgfplotsset{compat=1.18}

\usepackage{silence}
\usepackage{chngcntr}
\counterwithin*{equation}{section}
\counterwithin*{equation}{subsection}

\begin{document}
\ActivateWarningFilters[pdftoc]
\newcommand{\defeq}{:=}
\newcommand{\eps}{\varepsilon}

\newcommand{\liam}[1]{{\color{red} \textbf{Liam}: #1}} 
\newcommand{\avi}[1]{{\color{purple} \textbf{Avi}: #1}} 
\newcommand{\virgi}[1]{{\color{red} \textbf{Virgi}: #1}} 
\newcommand{\gur}[1]{{\color{blue} \textbf{Gur} #1}}
\newcommand{\claude}[1]{{\color{teal}\fontencoding{T1}\selectfont \textbf{Claude}: #1}}
\newcommand{\chatgpt}[1]{{\color{violet} \textbf{ChatGPT}: #1}}
\newcommand{\chatgt}[1]{\chatgpt{#1}}
\newcommand{\claudereply}[1]{{\color{orange}\fontencoding{T1}\selectfont \textbf{Claude-reply}: #1}}

\newcommand{\blue}[1]{{\color{blue}#1}}

\newcommand{\ReturnCode}{\textbf{return}}
\newcommand{\codestyle}[1]{\texttt{#1}}
\newcommand{\Initialize}{\mbox{\codestyle{Initialize}}}
\newcommand{\Cycle}{\mbox{\codestyle{Cycle}}}
\newcommand{\ADO}{\mbox{\codestyle{ADO}}}
\newcommand{\hADO}{\mbox{\codestyle{hADO}}}
\newcommand{\Query}{\mbox{\codestyle{.Query}}}
\newcommand{\THZQuery}{\mbox{\codestyle{TZQuery}}}
\newcommand{\ADOQuery}{\mbox{\codestyle{ADO.Query}}}
\newcommand{\ConstructADO}{\codestyle{ConstructADO}}
\newcommand{\FastAPSP}{\codestyle{FastAPSP}}
\newcommand{\Construct}{\mbox{\codestyle{Construct}}}
\newcommand{\Spanner}{\codestyle{Spanner}}
\newcommand{\Emulator}{\codestyle{Emulator}}
\newcommand{\RT}{\mbox{\codestyle{RT}}}
\renewcommand{\L}{\mbox{\codestyle{L}}}
\newcommand{\CycleOdd}{\codestyle{CycleOdd}}
\newcommand{\BallOrCycle}{\codestyle{BallOrCycle}}
\newcommand{\ClusterOrCycleBounded}{\codestyle{ClusterOrCycleBounded}}
\newcommand{\ApproximateANSC}{\codestyle{ApproximateANSC}}
\newcommand{\ApproximateANSCUnweighted}{\codestyle{ApproximateANSCUnweighted}}

\newcommand{\ClusterOrCycle}{\codestyle{ClusterOrCycle}}
\newcommand{\SimpleCycle}{\codestyle{SimpleCycle}}
\newcommand{\Next}{\codestyle{Next}}
\newcommand{\Sample}{\codestyle{Sample}}
\newcommand{\Dijkstra}{\codestyle{Dijkstra}}
\newcommand{\Preprocess}{\codestyle{Preprocess}}
\newcommand{\HashTable}{\codestyle{HashTable}}
\newcommand{\Heap}{\codestyle{Heap}}
\newcommand{\RelaxNext}{\codestyle{RelaxNext}}
\newcommand{\ODD}{_{\mathrm{ODD}}}

\newcommand{\PreprocessGraph}{\codestyle{Initialize}}
\newcommand{\Route}{\codestyle{Route}}
\newcommand{\TreeRoute}{\codestyle{TreeRoute}}
\newcommand{\N}{\mathbb{N}}
\newcommand{\MinCycle}{\codestyle{MinCycle}}
\newcommand{\Ball}{\codestyle{Ball}}
\newcommand{\DistanceOracle}{\codestyle{TZ-DistanceOracle}}
\newcommand{\SparseOrCycle}{\codestyle{SparseOrCycle}}
\newcommand{\Intersection}{\codestyle{Intersection}}
\newcommand{\CycleAdditive}{\codestyle{CycleAdditive}}
\newcommand{\GenerateSi}{\codestyle{ComputeS}}

\newcommand{\codeNull}{\codestyle{null}}
\newcommand{\codeYes}{\codestyle{Yes}}
\newcommand{\codeNo}{\codestyle{No}}
\newcommand{\codeAnd}{~ \mathrm{and} ~}
\newcommand{\codeOr}{~ \mathrm{or} ~}
\newcommand{\wt}{\ell}
\newcommand{\Cl}{CL}
\newcommand{\CL}{CL}
\newcommand{\cl}{c\ell}
\newcommand{\etal}{\textit{et~al.}}
\newcommand{\EQ}{\;=\;}
\newcommand{\GE}{\;\ge\;}
\newcommand{\Ot}{\tilde{O}}
\newcommand{\stactri}{\stackrel\triangle}
\newcommand{\sol}{\mathcal{Q}}

\newcommand{\tPSP}{$t$-PSP\xspace}
\newcommand{\nPSP}{$n$-PSP\xspace}
\newcommand{\mPSP}{$m$-PSP\xspace}
\newcommand{\ParameterizedNPSP}{\codestyle{\codestyle{Parameterized-nPSP}}}
\newcommand{\EE}{\mathbb{E}}
\newcommand{\RR}{\mathbb{R}}

\ifC
\renewcommand{\paragraph}[1]{\textbf{#1}}
\else
\fi

\DeclarePairedDelimiter{\ceil}{\lceil}{\rceil}
\DeclarePairedDelimiter{\floor}{\lfloor}{\rfloor}
\DeclarePairedDelimiter{\pair}{\langle}{\rangle}

\definecolor{lbentry}{RGB}{40,100,160}

\maketitle

\thispagestyle{empty}
\begin{abstract}
In this paper, we develop three new analytical tools for the Thorup--Zwick hierarchy: the \emph{Mixed-Level Growth}, \emph{Two-Level Growth}, and \emph{Long--Short Growth} lemmas. Using these tools, we obtain improved approximate distance oracles (ADOs), and algorithms for $n$-pairs shortest paths (\nPSP), multiple-source shortest paths (MSSP), and $ST$-shortest paths, as well as new hopsets.

One of our main results is a $(\frac 65 k+O(1))$-stretch ADO with $\Ot(mn^{1/k})$ preprocessing time and space (where $\Ot$ hides polylogarithmic factors), and $\Ot(n^{1/k})$ query time, breaking, for sparse graphs, the $(2k-O(1))$-stretch barrier for ADOs with $\Ot(n^{1+1/k})$ space and $\Ot(n^{1/k})$ query time. In unweighted graphs, we also obtain a $(\frac 65 k+O(1),O(k))$-approximate ADO with $\Ot(n^{1+1/k})$ space and $\Ot(n^{1/k})$ query time.

This also yields a $(\frac 65 k+O(1))$-approximation algorithm for $m$-pairs shortest paths (\mPSP) in $\Ot(mn^{1/k})$ time, answering the question of Chechik, Hoch, and Lifshitz [SODA'25] of whether the $2k-O(1)$ barrier for \nPSP in $\Ot(mn^{1/k})$ time can be broken.

For unweighted graphs, we give a new analysis of the existing $\Ot(mn^{1/k}+n^{1+2/k})$-time \nPSP algorithm of [Chechik et al., SODA'25], and show that, without any modification, it is a $(k,k-1)$-approximation, i.e., $d(u,v)\le\hat d(u,v)\le k\cdot d(u,v)+k-1$ for every queried pair. This improves its multiplicative factor from $\frac 43 k+O(1)$ to $k$, and the leading coefficient $k$ is tight for this algorithm. Moreover, for graphs with $m=\Omega(n^{1+1/k})$, this answers the question of Dalirrooyfard et al.~[FOCS'22] up to the additive $k-1$.

\end{abstract}
\clearpage
\newpage
\thispagestyle{empty}

\clearpage
\pagenumbering{arabic}
\setcounter{page}{1}
\section{Introduction}
\label{sec:introduction}
In their seminal work, Thorup and Zwick~\cite{DBLP:journals/jacm/ThorupZ05} introduced an approximate distance oracle (ADO) based on a hierarchical graph decomposition.
Let $G=(V,E)$ be an undirected graph with $n$ vertices and $m$ edges.
An ADO is a data structure that, given a pair of vertices $u,v\in V$, returns an estimate $\hat d(u,v)$ of the distance $d(u,v)$ between them. An ADO has stretch $\alpha$ if $d(u,v)\le \hat d(u,v)\le \alpha d(u,v)$ for every pair of vertices $u,v\in V$, and it is an $(\alpha,\beta)$-approximation if $d(u,v)\le \hat d(u,v)\le \alpha d(u,v)+\beta$.
For every integer $k\ge 1$, the Thorup--Zwick ADO has stretch $2k-1$, uses $\Ot(n^{1+1/k})$ space, answers queries in $O(k)$ time, and is constructed in $\Ot(mn^{1/k})$ time.\footnote{The $\Ot(\cdot)$ notation suppresses polylogarithmic factors in $n$. Since $n^{1/k}=O(1)$ for $k\ge\log n$, we assume that $k\le\log n$, so $\Ot(\cdot)$ also suppresses factors polynomial in $k$.}

The Thorup--Zwick hierarchy and related techniques have found applications in many graph algorithms and data structures, including
ADOs~\cite{RodittyTZ05,Nilsen12,WulffNilsen13,DBLP:conf/stoc/Chechik14,DBLP:conf/stoc/Chechik15,ElkinNW16,kadria2025new,kadria2025fasteralgorithms2k1stretchdistance,DBLP:conf/soda/ChechikZ22},
girth and shortest-cycle approximation~\cite{kadria2022girth,DBLP:conf/soda/HarbuzovaJWX24,DBLP:conf/soda/KadriaRSWZ23,kadria2026tightergirth,kadriadetgirth2026},
diameter approximation~\cite{CairoGR16,kirkpatrick2026diameter},
spanners and emulators~\cite{DBLP:conf/soda/ThorupZ06,DBLP:journals/rsa/BaswanaS07,Pettie09,DBLP:conf/soda/HarbuzovaJWX24},
APSP approximations~\cite{DBLP:journals/talg/Elkin05,DBLP:journals/siamcomp/BaswanaK10,AkavR20,DBLP:conf/soda/ChechikZ22,DBLP:conf/stoc/Roditty23,DoryFKNWV24,AkavR21,RodittyS25},
compact routing~\cite{DBLP:conf/spaa/ThorupZ01,DBLP:conf/podc/Chechik13,kadria2025routing},
distance labeling~\cite{DBLP:conf/wdag/AbrahamG11,SarmaDP15,ElkinFN18},
and hopsets~\cite{HuangP19,BenLevyP20,ElkinN19Linear,NeimanS22Hopsets}.

Our main contributions are three new analytical tools for the Thorup--Zwick hierarchy: the \emph{Mixed-Level Growth}, \emph{Two-Level Growth}, and \emph{Long--Short Growth} lemmas.
Using these tools, we obtain new algorithms for a variety of fundamental distance problems, improving upon the state of the art for each of them. We outline the main ideas behind these tools in \Cref{sec:technical-overview}.

Next, we describe our contributions for each problem and compare them with previous work. Throughout the paper, $k$ denotes a positive integer parameter. All of these results are direct applications of the new tools, and we believe that the tools can be used to improve algorithms for other graph problems as well.

\paragraph{ADOs with nonconstant query time.}
To improve upon the $(2k-1)$-stretch of the Thorup--Zwick ADO in weighted graphs, a common approach is to allow a larger query time and to focus on sparse graphs.\footnote{Assuming the Erd\H{o}s girth conjecture, the lower bound of
Thorup and Zwick~\cite{DBLP:journals/jacm/ThorupZ05} rules out stretch
below $2k-1$ with $O(n^{1+1/k})$ space in general graphs, regardless
of query time.}
ADOs with nonconstant query time have been studied in both weighted and
unweighted graphs~\cite{DBLP:conf/infocom/AgarwalGH11,PoratR13,AgarwalGH12,DBLP:conf/soda/AgarwalG13,PatrascuR14,DBLP:conf/esa/Agarwal14,DBLP:conf/focs/DalirrooyfardJW22,kadria2025new}.

We first discuss \emph{weighted} graphs.
Agarwal, Godfrey, and Har-Peled~\cite{DBLP:conf/infocom/AgarwalGH11} were the first to study this regime. They obtained a $(2k-3)$-stretch ADO with $O(m+n^{1+1/k})$ space and $O(\frac{m}{n}n^{1/k})$ query time, constructed in $\Ot(mn^{2/k})$ time. Later, Kadria and Roditty~\cite{kadria2025new} improved the stretch to $2k-5$ with the same space and query time, at the cost of an increased construction time of $\Ot(mn^{3/k})$.
These results left open whether the $2k-O(1)$ stretch barrier could be broken:
\begin{question}
\label{question:ado-break-2k}
    Is there an ADO for sparse weighted graphs (with $m=O(n)$) that uses $\Ot(n^{1+1/k})$ space and $\Ot(n^{1/k})$ query time, and has stretch $(2-\alpha)k$ for some constant $\alpha>0$?
\end{question}

The $2k-O(1)$ barrier has a common source. In the Thorup--Zwick query, every level at which the search fails to connect $u$ and $v$ allows the pivot distance to grow by up to $d(u,v)$. The estimate therefore grows by $2d(u,v)$ per level, which gives stretch $2k-1$ after $k$ levels.
All previous ADOs with $\Ot(n^{1/k})$ query time, as well as all previous $\Ot(mn^{1/k})$-time \nPSP algorithms, improve this analysis only at the first one or two levels. At every other level, they still pay the full $2d(u,v)$ of Thorup and Zwick~\cite{DBLP:journals/jacm/ThorupZ05}. Hence, their savings are additive, and their stretch remains $2k-O(1)$.
Our tools limit the growth of the pivot distances at \emph{every} level of the hierarchy. The savings thus accumulate over all $k$ levels, which reduces the leading coefficient of the stretch and breaks the $2k-O(1)$ barrier.

We answer \Cref{question:ado-break-2k} in the affirmative and break the $2k-O(1)$ stretch barrier by obtaining a $(\frac 65 k+O(1))$-stretch ADO with $\Ot(n^{1/k})$ query time, and $\Ot(mn^{1/k})$ space and preprocessing time.
\begin{theorem}[ADO-weighted]
\label{thm:intro-ado-weighted}\Copy{thm:intro-ado-weighted}{
    There is an $\Ot(mn^{1/k})$-time algorithm that constructs a $(\frac 65 k+7)$-stretch ADO for weighted undirected graphs, with $\Ot(mn^{1/k})$ space and $\Ot(n^{1/k})$ query time.
}
\end{theorem}

In \emph{unweighted} graphs, allowing an additive error removes the dependence on $m$
from the space and query time.
Here, the state of the art is the $(2k-5,6)$-approximate ADO of Kadria and Roditty~\cite{kadria2025new},
with $\Ot(n^{1+1/k})$ space, $O(n^{1/k})$ query time,
and $\Ot(mn^{3/k})$ preprocessing time.
The main question is whether the $2k-O(1)$ barrier can be broken:
\begin{question}
\label{question:ado-unweighted}
    Is there an ADO for unweighted graphs that uses $\Ot(n^{1+1/k})$ space and $\Ot(n^{1/k})$ query time, and is a $((2-\alpha)k,O(k))$-approximation for some constant $\alpha>0$?
\end{question}

Using our new tools, we answer \Cref{question:ado-unweighted} in the affirmative: we obtain a $(\frac 65 k+O(1), O(k))$-approximate ADO with $\Ot(n^{1/k})$ query time, $\Ot(n^{1+1/k})$ space, and $\Ot(mn^{1/k})$ preprocessing time.
\begin{corollary}[ADO-unweighted]
\label{thm:intro-ado-unweighted}\Copy{thm:intro-ado-unweighted}{
    There is an $\Ot(mn^{1/k})$-time algorithm that constructs a $(\frac 65 k+O(1),O(k))$-approximate ADO for unweighted undirected graphs, with $\Ot(n^{1+1/k})$ space and $\Ot(n^{1/k})$ query time.
}
\end{corollary}

\Cref{tab:oracle-comparison} compares our ADOs with the previous ADOs with nonconstant query time.

\begin{table}[t]
    \centering
    \small
    \renewcommand{\arraystretch}{1.15}
    \begin{tabular}{@{}llllll@{}}
        \toprule
        \textbf{Graphs} & \textbf{Reference} & \textbf{Preprocessing} & \textbf{Space} & \textbf{Query} & \textbf{Approximation} \\
        \midrule
        \multirow{5}{*}{Weighted}
        & \cite{DBLP:journals/jacm/ThorupZ05}
            & $\Ot(mn^{1/k})$ & $\Ot(n^{1+1/k})$ & $O(k)$ & $2k-1$ \\
        & \cite{DBLP:conf/focs/DalirrooyfardJW22}
            & $\Ot(mn^{1/k})$ & $\Ot(m+n^{1+1/k})$ & $O(\frac{m}{n} n^{1/k})$ & $2k-2$ \\
        & \cite{DBLP:conf/infocom/AgarwalGH11}
            & $\Ot(mn^{2/k})$ & $O(m+n^{1+1/k})$ & $O(\frac{m}{n} n^{1/k})$ & $2k-3$ \\
        & \cite{kadria2025new}
            & $\Ot(mn^{3/k})$ & $O(m+n^{1+1/k})$ & $O(\frac{m}{n} n^{1/k})$ & $2k-5$ \\
        & \Cref{thm:intro-ado-weighted}
            & $\Ot(mn^{1/k})$ & $\Ot(mn^{1/k})$ & $\Ot(n^{1/k})$ & $\frac 65 k+7$ \\
        \midrule
        \multirow{5}{*}{Unweighted}
        & \cite{DBLP:journals/jacm/ThorupZ05}
            & $\Ot(mn^{1/k})$ & $\Ot(n^{1+1/k})$ & $O(k)$ & $2k-1$ \\
        & \cite{DBLP:conf/focs/DalirrooyfardJW22}
            & $\Ot(mn^{1/k})$ & $\Ot(n^{1+1/k})$ & $O(n^{1/k})$ & $(2k-2,1)$ \\
        & \cite{DBLP:conf/infocom/AgarwalGH11}
            & $\Ot(mn^{2/k})$ & $\Ot(n^{1+1/k})$ & $O(n^{1/k})$ & $(2k-3,O(1))$ \\
        & \cite{kadria2025new}
            & $\Ot(mn^{3/k})$ & $\Ot(n^{1+1/k})$ & $O(n^{1/k})$ & $(2k-5,6)$ \\
        & \Cref{thm:intro-ado-unweighted}
            & $\Ot(mn^{1/k})$ & $\Ot(n^{1+1/k})$ & $\Ot(n^{1/k})$ & $(\frac 65 k+O(1),O(k))$ \\
        \bottomrule
    \end{tabular}
    \caption{ADOs with nonconstant query time.}
    \label{tab:oracle-comparison}
\end{table}

\paragraph{The \tPSP, \nPSP, and \mPSP problems.}
The $t$-pairs shortest paths problem (\tPSP) generalizes APSP
by asking only for the distances between a prescribed set of $t$ pairs, instead of all $n^2$ pairs.
Specifically, given a set $I\subseteq V\times V$ of $t$ pairs,
the goal is to compute an estimate $\hat d(u,v)$ for every
$(u,v)\in I$. 
The \tPSP problem has been studied since the 1990s by Cohen~\cite{DBLP:journals/siamcomp/Cohen98} and by Aingworth,
Chekuri, Indyk, and Motwani~\cite{ACIM99}, and has received renewed attention in recent years~\cite{DBLP:conf/focs/DalirrooyfardJW22,DBLP:conf/soda/ChechikHL25,kadria2026improved}.
This line of research has focused primarily on the case $t=n$, which we denote by \nPSP.
We write \mPSP for the case $t=m$.

We first discuss \nPSP algorithms running in $\Ot(mn^{1/k})$ time, and then those with an additional $n^{1+2/k}$ term in the running time.

\paragraph{Running time $\Ot(mn^{1/k})$.}
For \emph{unweighted} graphs, Dalirrooyfard, Jin,
Vassilevska Williams, and Wein~\cite{DBLP:conf/focs/DalirrooyfardJW22} obtained a
$(2k-2,1)$-approximation for \nPSP in $\Ot(mn^{1/k})$ time. Chechik, Hoch, and Lifshitz~\cite{DBLP:conf/soda/ChechikHL25} later obtained a purely multiplicative $(2k-3)$-approximation in the same running time, for $k\ge 3$, in both unweighted and weighted graphs.
The main open problem in this regime is whether the $2k-O(1)$ barrier can be broken:
\begin{question}
\label{question:break-2k}
    Is there a $(2-\alpha)k$-approximation algorithm for \nPSP in unweighted graphs that runs in $\Ot(mn^{1/k})$ time, for some constant $\alpha>0$? Is there such an algorithm for weighted graphs?
\end{question}
We answer both parts of \Cref{question:break-2k} in the affirmative. We present a $(\frac 65 k+O(1))$-approximation algorithm that runs in $\Ot(mn^{1/k})$ time on weighted graphs, and hence also on unweighted graphs. Moreover, it handles $m$ pairs instead of $n$.

\begin{theorem}[\mPSP-weighted]\label{thm:intro-weighted}\Copy{thm:intro-weighted}{
    There is an $\Ot(mn^{1/k})$-time $(\frac 65 k+7)$-approximation algorithm for \mPSP in weighted undirected graphs.
}
\end{theorem}

\paragraph{Running time $\Ot(mn^{1/k}+n^{1+2/k})$.}
With an additional $n^{1+2/k}$ term in the running time, better leading coefficients were known.
In unweighted graphs, Chechik, Hoch, and Lifshitz~\cite{DBLP:conf/soda/ChechikHL25} obtained a
$(\lceil4k/3\rceil-1,\lceil4k/3\rceil-1)$-approximation, and Kadria, Roditty, and Vassilevska
Williams~\cite{kadria2026improved} improved it to a
purely multiplicative $\lceil4k/3-5/3\rceil$-approximation, for $k\ge 4$. Both results analyze the same algorithm, which connects $u$ and $v$ through the \emph{middle vertex} of their shortest path (see \Cref{sec:technical-overview}).
In weighted graphs, Kadria, Roditty, and Vassilevska Williams~\cite{kadria2026improved} obtained a $1.622k$-approximation. \Cref{thm:intro-weighted} already improves upon this result. It has a smaller leading coefficient, it runs in $\Ot(mn^{1/k})$ time, and it handles $m$ pairs instead of $n$.
The main open problem in this regime is whether this running time suffices for stretch $k$:
\begin{question}
\label{question:k-approx}
Is there a multiplicative $k$-approximation algorithm for \nPSP that runs in $\Ot(mn^{1/k}+n^{1+2/k})$ time?
\end{question}

In unweighted graphs, we answer \Cref{question:k-approx} in the affirmative. 
Using our new analytical tools, we show that the algorithm of~\cite{kadria2026improved} is a $(k,k-1)$-approximation algorithm for \nPSP.
This improves the leading coefficient from $\frac 43k$ to exactly $k$ with the same running time.
Moreover, the leading coefficient $k$ is tight for this algorithm (see \Cref{sec:k-stretch}).
\begin{theorem}[\nPSP-unweighted]\label{thm:intro-unweighted-npsp}\Copy{thm:intro-unweighted-npsp}{
    There is an $\Ot(mn^{1/k}+n^{1+2/k})$-time $(k,k-1)$-approximation algorithm for \nPSP in unweighted undirected graphs.
}
\end{theorem}

When $m=\Omega(n^{1+1/k})$, the running time of \Cref{thm:intro-unweighted-npsp} is $\Ot(mn^{1/k})$. Hence, for such graphs, \Cref{thm:intro-unweighted-npsp} also answers, up to the same additive $k-1$, the question of Dalirrooyfard, Jin, Vassilevska Williams, and Wein~\cite{DBLP:conf/focs/DalirrooyfardJW22}, who asked whether a $k$-approximation for \nPSP is possible in $\Ot(mn^{1/k})$ time.

\Cref{tab:main-comparison} summarizes our \nPSP and \mPSP results and compares them with previous algorithms.

\begin{table}[t]
    \centering
    \small
    \renewcommand{\arraystretch}{1.15}
    \begin{tabular}{@{}lllll@{}}
        \toprule
        \textbf{Graphs} & \textbf{Reference} & \textbf{\#Pairs} & \textbf{Running time} & \textbf{Approximation} \\
        \midrule
        \multirow{4}{*}{Weighted}
        & \cite{DBLP:journals/jacm/ThorupZ05}
            & $t$ & $\Ot(mn^{1/k}+t)$ & $2k-1$ \\
        & \cite{DBLP:conf/soda/ChechikHL25}
            & $n$ & $\Ot(mn^{1/k})$ & $2k-3$ \\
        & \cite{kadria2026improved}
            & $n$ & $\Ot(mn^{1/k}+n^{1+2/k})$ & $1.622k$ \\
        & \Cref{thm:intro-weighted}
            & $m$ & $\Ot(mn^{1/k})$ & $\frac 65 k+7$ \\
        \midrule
        \multirow{6}{*}{Unweighted}
        & \cite{DBLP:conf/focs/DalirrooyfardJW22}
            & $n$ & $\Ot(mn^{1/k})$ & $(2k-2,1)$ \\
        & \cite{DBLP:conf/soda/ChechikHL25}
            & $n$ & $\Ot(mn^{1/k})$ & $2k-3$ \\
        & \cite{DBLP:conf/soda/ChechikHL25}
            & $n$ & $\Ot(mn^{1/k}+n^{1+2/k})$
            & $(\lceil4k/3\rceil-1,\lceil4k/3\rceil-1)$ \\
        & \cite{kadria2026improved}
            & $n$ & $\Ot(mn^{1/k}+n^{1+2/k})$ & $\lceil4k/3-5/3\rceil$ \\
        & \Cref{thm:intro-unweighted-npsp}
            & $n$ & $\Ot(mn^{1/k}+n^{1+2/k})$ & $(k,k-1)$ \\
        & \Cref{thm:intro-weighted}
            & $m$ & $\Ot(mn^{1/k})$ & $\frac 65 k+7$ \\
        \bottomrule
    \end{tabular}
    \caption{\nPSP and \mPSP algorithms.}
    \label{tab:main-comparison}
\end{table}

\paragraph{$(t,s)$-PSP: $ST$-shortest paths and MSSP.}
Two extensively studied problems that are closely related to \tPSP (see, e.g.,~\cite{klein2005multiple,cabello2007multiple,DBLP:journals/siamcomp/CabelloCE13,DBLP:conf/stoc/EisenstatK13,DBLP:journals/siamcomp/DorHZ00,DBLP:journals/talg/Elkin05,DBLP:conf/focs/DalirrooyfardJW22}) are multiple-source shortest paths (MSSP) and $ST$-shortest paths ($ST$-SP).
Given a set $S\subseteq V$, MSSP asks to estimate $d(u,v)$ for every $(u,v)\in S\times V$.
Given two sets $S,T\subseteq V$, $ST$-SP asks to estimate $d(u,v)$ for every $(u,v)\in S\times T$.
All of these problems are special cases of a single source-sensitive problem, which we call \emph{$(t,s)$-PSP}.
\begin{definition}[$(t,s)$-PSP]
The input is a set
$I$ of $t$ prescribed pairs such that there exists a set $A\subseteq V$ of $s$
\emph{sources} that hits every pair, that is, $A\cap\{u,v\}\neq\emptyset$ for
every $\pair{u,v}\in I$. The goal is to compute an estimate $\hat{d}(u,v)$ for every $\pair{u,v}\in I$.\footnote{The set $A$ does not need to be given. Scanning the pairs one by one and adding both endpoints of every pair that is not yet hit gives, in $O(t)$ time, a hitting set of size at most $2s$. }
\end{definition}

For example, \nPSP is $(n,n)$-PSP, \mPSP is $(m,n)$-PSP, and \tPSP is $(t,\min\{t,n\})$-PSP.
Similarly, $ST$-SP is $(|S|\cdot|T|,\min\{|S|,|T|\})$-PSP, and MSSP is $(n\cdot|S|,|S|)$-PSP.

Using our new tools and ideas from the parameterized distance oracles of~\cite{kadria2025fasteralgorithms2k1stretchdistance, RodittyTZ05}, we obtain the following result for $(t,s)$-PSP in weighted graphs:
\begin{theorem}[$(t,s)$-PSP-weighted]
\label{thm:intro-tsPSP-weighted}\Copy{thm:intro-tsPSP-weighted}{
    There is an $\Ot((m+t)s^{1/k})$-time $(\frac 65 k+11)$-approximation algorithm for $(t,s)$-PSP in weighted undirected graphs.
}
\end{theorem}

Thus, the $n^{1/k}$ factor in the running time of \Cref{thm:intro-weighted} is replaced by $s^{1/k}$: the cost is governed by the number of \emph{sources} $s$, rather than by the number of vertices $n$.\footnote{The constant term in the stretch grows from $7$ in \Cref{thm:intro-weighted} to $11$, since with $s<n$ sources the hierarchy no longer starts at $h_0(u)=h_0(v)=0$, which adds at most $4d(u,v)$ to the estimate. See \Cref{sec:ts-psp}.}

For unweighted graphs, we obtain:
\begin{theorem}[$(t,s)$-PSP-unweighted]
\label{thm:intro-tsPSP-unweighted}\Copy{thm:intro-tsPSP-unweighted}{
    There is an $\Ot(ms^{1/k}+(n+t)s^{2/k})$-time $(k+2,k-1)$-approximation algorithm for $(t,s)$-PSP in unweighted undirected graphs.
}
\end{theorem}

As immediate corollaries, we obtain algorithms for $ST$-SP and MSSP.
Let $\sigma=\min\{|S|,|T|\}$. For $ST$-SP, we substitute $s=\sigma$ and $t=|S|\cdot|T|$:
\begin{corollary}[$ST$-SP-weighted]
\label{cor:intro-STSP-weighted}
There is an $\Ot((m+|S|\cdot|T|)\sigma^{1/k})$-time $(\frac 65 k+11)$-approximation algorithm for $ST$-SP in weighted undirected graphs.
\end{corollary}

\begin{corollary}[$ST$-SP-unweighted]
\label{cor:intro-STSP-unweighted}
There is an $\Ot(m\sigma^{1/k}+(n+|S|\cdot|T|)\sigma^{2/k})$-time $(k+2,k-1)$-approximation algorithm for $ST$-SP in unweighted undirected graphs.
\end{corollary}

For MSSP, we substitute $s=|S|$ and $t=n|S|$:
\begin{corollary}[MSSP-weighted]\label{cor:MSSP-weighted}
There is an $\Ot((m+n|S|)|S|^{1/k})$-time $(\frac 65 k+11)$-approximation algorithm for MSSP in weighted undirected graphs.
\end{corollary}

\begin{corollary}[MSSP-unweighted]\label{cor:MSSP-unweighted}
There is an $\Ot(m|S|^{1/k}+n|S|^{1+2/k})$-time $(k+2,k-1)$-approximation algorithm for MSSP in unweighted undirected graphs.
\end{corollary}

\Cref{tab:ts-comparison} compares these results with the source-restricted ADO of Roditty, Thorup, and Zwick~\cite{RodittyTZ05}.

\begin{table}[t]
    \centering
    \small
    \renewcommand{\arraystretch}{1.15}
    \begin{tabular}{@{}lllll@{}}
        \toprule
        \textbf{Graphs} & \textbf{Problem} & \textbf{Reference} & \textbf{Running time} & \textbf{Approximation} \\
        \midrule
        \multirow{4}{*}{Weighted}
        & \multirow{2}{*}{$(t,s)$-PSP}
            & \cite{RodittyTZ05} & $\Ot(ms^{1/k}+t)$ & $2k-1$ \\
        & & \Cref{thm:intro-tsPSP-weighted} & $\Ot(ms^{1/k}+t\cdot s^{1/k})$ & $\frac 65 k+11$ \\
        \cmidrule(l){2-5}
        & \multirow{2}{*}{MSSP}
            & \cite{RodittyTZ05} & $\Ot(ms^{1/k}+ns)$ & $2k-1$ \\
        & & \Cref{cor:MSSP-weighted} & $\Ot(ms^{1/k}+ns\cdot s^{1/k})$ & $\frac 65 k+11$ \\
        \midrule
        \multirow{4}{*}{Unweighted}
        & \multirow{2}{*}{$(t,s)$-PSP}
            & \cite{RodittyTZ05} & $\Ot(ms^{1/k}+t)$ & $2k-1$ \\
        & & \Cref{thm:intro-tsPSP-unweighted} & $\Ot(ms^{1/k}+ns^{2/k}+t\cdot s^{2/k})$ & $(k+2,k-1)$ \\
        \cmidrule(l){2-5}
        & \multirow{2}{*}{MSSP}
            & \cite{RodittyTZ05} & $\Ot(ms^{1/k}+ns)$ & $2k-1$ \\
        & & \Cref{cor:MSSP-unweighted} & $\Ot(ms^{1/k}+ns\cdot s^{2/k})$ & $(k+2,k-1)$ \\
        \bottomrule
    \end{tabular}
    \caption{$(t,s)$-PSP and MSSP algorithms. Here, $s$ is the number of sources (for MSSP, $s=|S|$).}
    \label{tab:ts-comparison}
\end{table}

Finally, by taking $s=\min\{t,n\}$ in \Cref{thm:intro-tsPSP-weighted,thm:intro-tsPSP-unweighted}, we obtain the following algorithms for \tPSP.
When $t\ll n$, they improve upon \Cref{thm:intro-weighted,thm:intro-unweighted-npsp} by replacing the $n^{1/k}$ factor in the running time with $t^{1/k}$.
In the unweighted case, this comes with a slightly larger stretch, $(k+2,k-1)$ instead of $(k,k-1)$.

\begin{corollary}[\tPSP-weighted]
\label{cor:intro-tPSP-weighted}
There is an $\Ot((m+t)t^{1/k})$-time $(\frac 65 k+11)$-approximation algorithm for \tPSP in weighted undirected graphs.
\end{corollary}

\begin{corollary}[\tPSP-unweighted]
\label{cor:intro-tPSP-unweighted}
There is an $\Ot(mt^{1/k}+(n+t)t^{2/k})$-time $(k+2,k-1)$-approximation algorithm for \tPSP in unweighted undirected graphs.
\end{corollary}

\paragraph{Hopsets.}
A $b$-hops $(\alpha,\beta)$-approximation hopset is a set $\mathcal{H}$ of weighted edges whose addition to $G$ preserves distances and gives every pair $u,v\in V$ a path in $G\cup\mathcal{H}$ with at most $b$ edges and length at most $\alpha d(u,v)+\beta$. Hopsets have been studied extensively, both in upper bounds~\cite{Cohen00,MillerPVX15Hopsets,HenzingerKN18Hopsets,ElkinN19Hopsets,HuangP19,BenLevyP20,ElkinM21Hopsets,ElkinN19Linear,KoganP22Directed,NeimanS22Hopsets,BernsteinW23Hopsets,AshvinkumarBDGW25Hopsets} and in lower bounds~\cite{DBLP:journals/siamcomp/AbboudBP18,DBLP:conf/focs/KoganP22,NeimanS22Hopsets,BodwinH23Hopsets,HoppenworthXX25Hopsets}.

Huang and Pettie~\cite{HuangP19} observed that the Thorup--Zwick~\cite{DBLP:journals/jacm/ThorupZ05} distance oracle yields a $2$-hops $(2k-1,0)$-approximation hopset of size $\Ot(n^{1+1/k})$, and asked the following question:
\begin{question}
\label{question:constant-hopbound}
Are other tradeoffs between size and stretch possible for hopsets whose hopbound is a fixed constant, such as $3$ or $4$, independent of $k$?
\end{question}

We answer this question in the affirmative and present several new hopsets with constant hopbounds that break the $2k-O(1)$ barrier. For unweighted graphs, we obtain a $(k,k-1)$-approximation with $4$ hops:
\begin{corollary}[Hopset-unweighted]
\label{cor:intro-hopset-unweighted}\Copy{cor:intro-hopset-unweighted}{
    There is a $4$-hops $(k,k-1)$-approximation hopset of size $\Ot(n^{1+1/k})$ for unweighted graphs.
}
\end{corollary}

For weighted graphs, we obtain three hopsets with $3$, $4$, and $5$ hops. Each additional hop improves the tradeoff: the fourth hop reduces stretch, and the fifth reduces size. First, with $3$ hops and $\Ot(mn^{1/k})$ edges, we obtain stretch $\frac 65 k+7$:

\begin{corollary}[Hopset-weighted-3-hops]
\label{cor:intro-hopset-weighted-3hop}\Copy{cor:intro-hopset-weighted-3hop}{
    There is a $3$-hops $(\frac 65 k+7)$-stretch hopset of size $\Ot(mn^{1/k})$ for weighted graphs.
}
\end{corollary}

With one additional hop, we reduce the stretch to $k$ at the same size.
\begin{corollary}[Hopset-weighted-4-hops]
\label{cor:intro-hopset-weighted-4hop}\Copy{cor:intro-hopset-weighted-4hop}{
    There is a $4$-hops $k$-stretch hopset of size $\Ot(mn^{1/k})$ for weighted graphs.
}
\end{corollary}

A fifth hop preserves stretch $k$ while reducing the hopset size to $\Ot(n^{1+1/k})$.
\begin{corollary}[Hopset-weighted]
\label{cor:intro-hopset-weighted}\Copy{cor:intro-hopset-weighted}{
    There is a $5$-hops $k$-stretch hopset of size $\Ot(n^{1+1/k})$ for weighted graphs.
}
\end{corollary}

The Thorup--Zwick bunches and pivots used by \Cref{thm:intro-unweighted-npsp,cor:intro-hopset-unweighted} are also nearly optimal in the pivot-based model of Kadria and Roditty~\cite{KadriaR26pivot}. In a $p$-pivot $\beta$-hop ADO, every vertex has at most $p$ pivots, including itself, and every estimate is the length of a walk of stored distances with at most $\beta$ hops, where, unlike in a hopset, pivot jumps between a vertex and one of its pivots are not counted as hops.
Every estimate in our analysis of \Cref{thm:intro-unweighted-npsp} is the length of a walk $u\to p_u\to p_x\to x\to p'_x\to p_v\to v$, where $p_u$, $p_x$, $p'_x$, and $p_v$ are Thorup--Zwick pivots of $u$, $x$, $x$, and $v$, respectively, and the only hops are the bunch distances from $p_u$ to $p_x$ and from $p'_x$ to $p_v$.
Since this holds for every pair $u,v\in V$, the Thorup--Zwick bunch and pivot distances form a $k$-pivot $2$-hop $(k,k-1)$-approximation ADO with $\Ot(n^{1+1/k})$ space.
Under the Erd\H{o}s girth conjecture, there are sparse unweighted graphs on which every $\Ot(1)$-pivot $2$-hop ADO with $\Ot(n^{1+1/k})$ space has stretch at least $k-1$, even with an additive error independent of $n$~\cite{KadriaR26pivot}.
Thus, within this model, the leading coefficient $k$ is optimal up to an additive $1$.

\paragraph{Short paths.}
In addition, we obtain an improved algorithm for \mPSP when the shortest paths of the queried pairs have at most two edges. This result also answers a question of Kadria and Roditty~\cite{kadria2025new}, who asked whether stretch exactly $k$ is achievable for vertex pairs at distance $2$ in unweighted graphs:
\begin{theorem}[\mPSP-short paths]\label{thm:nPSP-2-hop}\Copy{thm:nPSP-2-hop}{
    There is an $\Ot(mn^{1/k})$-time $k$-approximation algorithm for \mPSP in weighted undirected graphs, for queried pairs $(u,v)$ whose shortest path $P(u,v)$ has at most $2$ edges.
}
\end{theorem}



\paragraph{Related work: Conditional lower bounds.}
Abboud, Bringmann, Khoury, and Zamir~\cite{DBLP:conf/stoc/AbboudBKZ22} initiated the study of conditional lower bounds for approximate distance oracles with nonconstant query time. Building on their work, Abboud, Bringmann, and Fischer~\cite{DBLP:conf/stoc/AbboudBF23} and, independently, Jin and Xu~\cite{jin2023removing} proved stronger lower bounds under the 3SUM hypothesis: for a fixed $k$, every ADO with $\Ot(mn^{1/k})$ preprocessing time and $n^{o(1)}$ query time must have stretch at least $k-O(1)$, even in unweighted graphs.
Jin and Xu~\cite{jin2023removing} also proved lower bounds for the offline setting, in which all pairs are given in advance. In particular, their result~\cite[Theorem~1.9]{jin2023removing} implies that, for every integer $k\ge 3$ and constant $\varepsilon>0$, there is no $(k/2-\varepsilon)$-approximation algorithm for \mPSP that runs in $\Ot(mn^{1/k})$ time, even in unweighted graphs, unless the 3SUM hypothesis fails. Hence, the leading coefficient $\frac 65$ of \Cref{thm:intro-weighted} is within a factor of $\frac{12}{5}$ of the best possible.

For pairs at distance $2$, Jin and Xu~\cite{jin2023removing} proved, under the 3SUM hypothesis, that for every constant integer $k\ge 2$ and every $\varepsilon,\delta>0$, there is no $O(m^{1+\frac{1}{2k-1}-\varepsilon})$-time algorithm that $(k-\delta)$-approximates the distances of $m$ given pairs in an unweighted undirected graph with $m=\Theta(n^{1+\frac{1}{2k-2}})$, even when every queried pair is at distance at most $2$.
On such graphs, $m^{\frac{1}{2k-1}}=\Theta(n^{\frac{1}{2k-2}})$, so their lower bound rules out stretch $k-\delta$ in $O(mn^{\frac{1}{2k-2}-\varepsilon})$ time.
Applying it with $\frac k2+1$ in place of $k$ (for even $k$) shows that no algorithm with stretch $\frac k2+1-\delta$ runs in $O(mn^{\frac 1k-\varepsilon})$ time.
Since a pair at distance $2$ in an unweighted graph has a shortest path with $2$ edges, \Cref{thm:nPSP-2-hop} applies, and it achieves stretch $k$ in $\Ot(mn^{1/k})$ time, which is within a factor of $2$ of this lower bound.
A similar factor of $2$ in the stretch appears for ADOs. Jin and Xu~\cite{jin2023removing} also rule out a $(k-\delta)$-stretch ADO with $O(n^{1+\frac{1}{k-1}-\varepsilon})$ preprocessing time and $n^{o(1)}$ query time on graphs with $O(n)$ edges.
Applying it with $k+1$ in place of $k$ shows that no such ADO with stretch $k+1-\delta$ has $O(n^{1+\frac 1k-\varepsilon})$ preprocessing time, whereas the Thorup--Zwick ADO with stretch $2k-1$ is constructed in $\Ot(n^{1+1/k})$ time on such graphs.
It is therefore plausible that in both cases the lower bound, rather than the algorithm, can be improved, in which case \Cref{thm:nPSP-2-hop} would be tight.

\paragraph{Organization.}
In \Cref{sec:prelim} we fix our notation and review the Thorup--Zwick hierarchy, and in \Cref{sec:technical-overview} we give a technical overview of our three growth lemmas and of how they replace the middle-vertex analysis.
In \Cref{sec:k-stretch} we prove the Mixed-Level Growth Lemma and use it to obtain the unweighted $(k,k-1)$-approximation of \Cref{thm:intro-unweighted-npsp}, and in \Cref{sec:65k} we present the $(\frac 65 k+O(1))$-approximation algorithm of \Cref{thm:intro-weighted}, together with the Two-Level Growth Lemma and the Long--Short Growth Lemma.
In \Cref{sec:short-paths} we prove \Cref{thm:nPSP-2-hop}, which shows that stretch $k$ is achievable for pairs whose shortest path has at most two edges.
In \Cref{sec:dos} we derive our ADOs, \Cref{thm:intro-ado-weighted,thm:intro-ado-unweighted}, and our hopsets, and in \Cref{sec:ts-psp} we generalize our algorithms to $(t,s)$-PSP, obtaining \Cref{thm:intro-tsPSP-weighted,thm:intro-tsPSP-unweighted} and their corollaries for $ST$-SP, MSSP, and \tPSP.
Finally, the numerical optimization lemmas used by our stretch analyses, \Cref{lem:k-optimization,lem:six-fifths-optimization}, are proved in \Cref{app:numerical-lemma}.

\section{Preliminaries}
\label{sec:prelim}
Let $G=(V,E)$ be an undirected graph with $n=|V|$ vertices and $m=|E|$ edges. Throughout the paper, we consider both unweighted graphs and weighted graphs with nonnegative real edge weights. We denote the weight of an edge $(x,y)\in E$ by $\wt(x,y)$, and $\wt(x,y)=1$ in unweighted graphs. Each edge $\{x,y\}\in E$ has two orientations, $(x,y)$ and $(y,x)$, and we refer to them as oriented edges. We assume that $G$ is connected, so all distances are finite. 
Let $u,v\in V$. The distance $d(u,v)$ between $u$ and $v$ is the length of a shortest path between $u$ and $v$. 
Let $P(u,v)$ be a shortest path between $u$ and $v$. For a subset $X \subseteq V$ and a vertex $v \in V$, we set $d(v,X) = \min_{x \in X} d(v,x)$.

\paragraph{The Thorup--Zwick hierarchy.} In our algorithms, as well as in many distance-related algorithms (\nPSP, distance oracles, all-pairs shortest paths, multiple-source shortest paths, spanners, etc.), one uses a vertex hierarchy $V=A_0\supseteq A_1 \supseteq \ldots \supseteq A_k=\emptyset$, where $A_{i+1}$ contains every vertex from $A_i$ with probability $n^{-1/k}$ for every $0 \le i < k-1$.
For every vertex $u \in V$, let $p_i(u) \in \arg \min_{w \in A_i} d(u,w)$, $h_i(u):=d(u,p_i(u))$, and
\[
B_i(u) = \{ w \in A_{i} \mid d(u,w) < h_{i+1}(u) \},
\]
and let $B(u)=\cup_i B_i(u)$. We call $p_i(u)$ the $i$-th pivot of $u$, and $B(u)$ the bunch of $u$. Note that $u\in B(u)$ for every $u\in V$.
Since $A_k=\emptyset$, we have $h_k(u)=\infty$, and hence $B_{k-1}(u)=A_{k-1}$. We write $[k]=\{0,1,\ldots,k-1\}$ for the set of levels.
For a vertex $x$, we write $\Delta_i(x)=h_i(x)-h_{i-1}(x)$ for $1\le i<k$ and $\nabla_i(x)=h_i(x)-h_{i-2}(x)=\Delta_i(x)+\Delta_{i-1}(x)$ for $2\le i<k$, and call them the one-level and two-level growth of $x$ at level $i$.

Thorup and Zwick~\cite{DBLP:journals/jacm/ThorupZ05} also presented an efficient algorithm for computing $B(u)$ for every $u\in V$, as described in the following lemma.
\begin{lemma}[\cite{DBLP:journals/jacm/ThorupZ05}]\label{lem:tz-construction}
    There is an $\Ot(mn^{1/k})$-time algorithm that computes, for every $u\in V$, the set $B(u)$ and the distances from $u$ to every $w\in B(u)$.
\end{lemma}

Thorup and Zwick~\cite{DBLP:journals/jacm/ThorupZ05} showed that this hierarchy can be used as an approximate distance oracle. The following lemma is the main ingredient in their analysis. We present the proof as a warm-up for later sections.
\begin{lemma}[\cite{DBLP:journals/jacm/ThorupZ05}]
\label{lem:tz-bunch}
Let \(u,v\in V\) and \(1\le i\le k-1\). If
\(
d(u,v)+h_{i-1}(v)<h_i(u),
\)
then
\(
p_{i-1}(v)\in B(u),
\)
and
\[
d(u,p_{i-1}(v)) + d(p_{i-1}(v),v) \le 2h_i(u) - d(u,v).
\]
\end{lemma}

\begin{proof}
By the triangle inequality,
\(
d\bigl(u,p_{i-1}(v)\bigr)
\le d(u,v)+h_{i-1}(v)
<h_i(u).
\)
Since \(p_{i-1}(v)\in A_{i-1}\) and \(h_i(u)=d(u,A_i)\), it follows that
\(
p_{i-1}(v)\in B_{i-1}(u)\subseteq B(u).
\)

In addition, 
\[
d(u,p_{i-1}(v)) + d(p_{i-1}(v),v) \leq h_i(u)+h_{i-1}(v)<2h_i(u)-d(u,v),
\]
as required.
\end{proof}

Moreover, a one-level gap of more than twice the distance gives both inclusions at once.
\begin{lemma}\label{lem:one-level-gap}
Let \(x,v\in V\) and \(1\le j\le k-1\). If
\(
h_j(v)-h_{j-1}(v)>2d(x,v),
\)
then \(p_{j-1}(x)\in B(v)\) and \(p_{j-1}(v)\in B(x)\).
\end{lemma}

\begin{proof}
Since \(h_{j-1}(x)\le h_{j-1}(v)+d(x,v)\), we have
\(
d(v,x)+h_{j-1}(x)\le 2d(x,v)+h_{j-1}(v)<h_j(v),
\)
and \Cref{lem:tz-bunch} gives \(p_{j-1}(x)\in B(v)\).
Since \(h_j(x)\ge h_j(v)-d(x,v)\), we have
\(
d(x,v)+h_{j-1}(v)<h_j(v)-d(x,v)\le h_j(x),
\)
and \Cref{lem:tz-bunch} gives \(p_{j-1}(v)\in B(x)\).
\end{proof}

Thorup and Zwick~\cite{DBLP:journals/jacm/ThorupZ05} proved the following bound on the size of the bunches. 
\begin{lemma}[Lemma~3.5 in~\cite{DBLP:journals/jacm/ThorupZ05}]
    $|B(u)| = O(k\cdot n^{1/k}\log^{1-1/k} n)=\Ot(n^{1/k})$ with high probability.
\end{lemma}

\section{Technical overview}
\label{sec:technical-overview}
In this section, we give an overview of the new machinery developed in this paper, specifically the Mixed-Level Growth, Two-Level Growth, and Long--Short Growth lemmas. For simplicity, we present their application to the \nPSP and \mPSP problems, and then briefly discuss the implications for other problems.

We begin by explaining how previous analyses improve upon the classical $2k-1$ stretch, and what limits their improvement. 
The baseline is to construct the Thorup--Zwick oracle and query it: the query searches for a pivot
$p_i(u)\in B(v)$ or $p_i(v)\in B(u)$ through which to connect $u$ and $v$.
This gives stretch $2k-1$ with $O(k)$ query time and $\Ot(mn^{1/k})$ preprocessing time, and hence an $\Ot(mn^{1/k}+t)$-time algorithm for \tPSP with the same stretch.
For \nPSP and \mPSP, we can inspect more than the $O(k)$ pivot-to-bunch candidates of a standard oracle query. These additional candidates allow us to improve on the $2k-1$ stretch.

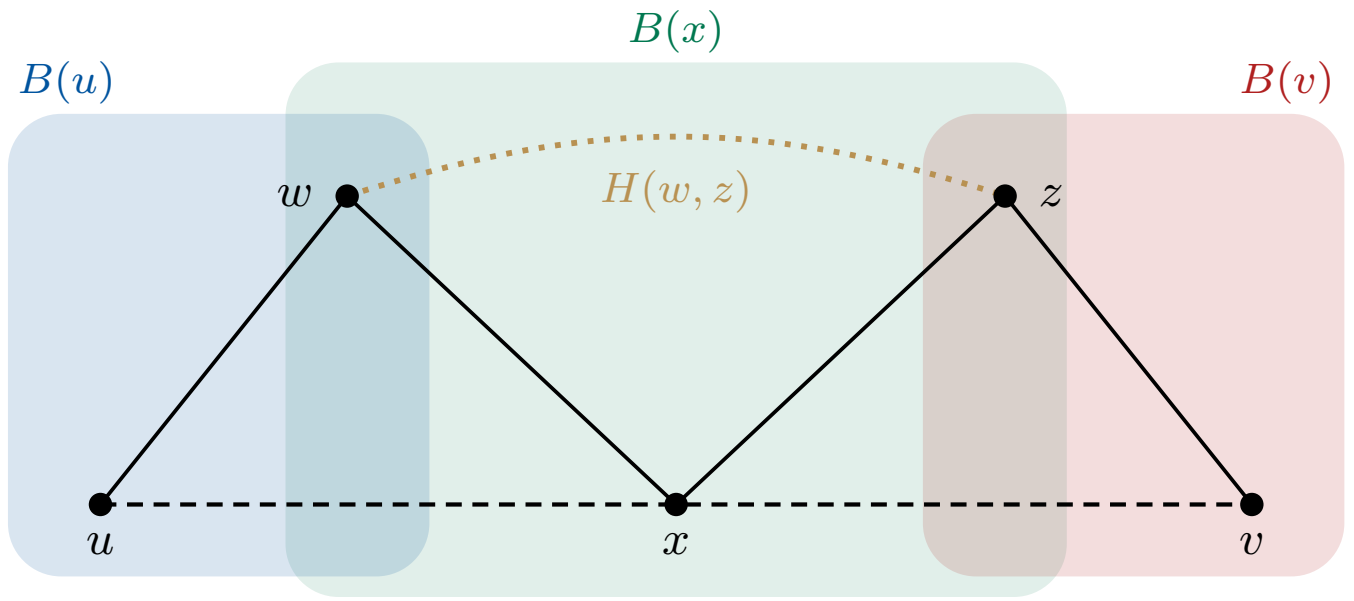
\begin{figure}[t]
    \centering
    \tikzset{every picture/.style={line width=0.75pt}} 
    \definecolor{bunchu}{RGB}{0,84,159}
    \definecolor{bunchx}{RGB}{0,120,80}
    \definecolor{bunchv}{RGB}{176,32,32}
    \definecolor{hsand}{RGB}{184,145,82}
\begin{tikzpicture}[x=0.68cm,y=0.68cm,line width=0.75pt,
    every node/.style={font=\small,inner sep=1.5pt},
    leg/.style={line width=0.7pt}]

  \coordinate (u) at (0,0);
  \coordinate (w) at (2.4,3.0);
  \coordinate (x) at (5.6,0);
  \coordinate (z) at (8.8,3.0);
  \coordinate (v) at (11.2,0);

  \begin{scope}[rounded corners=10pt]
    \fill[bunchx,fill opacity=0.12] (1.8,-0.9) rectangle (9.4,4.3);
    \fill[bunchu,fill opacity=0.15] (-0.9,-0.7) rectangle (3.2,3.8);
    \fill[bunchv,fill opacity=0.15] (8.0,-0.7) rectangle (12.1,3.8);
  \end{scope}
  \node[text=bunchu,font=\footnotesize,anchor=south west] at (-0.9,3.8) {$B(u)$};
  \node[text=bunchx,font=\footnotesize,anchor=south] at (5.6,4.3) {$B(x)$};
  \node[text=bunchv,font=\footnotesize,anchor=south east] at (12.1,3.8) {$B(v)$};

  \draw[line width=0.8pt,densely dashed] (u) -- (x) -- (v);

  \draw[leg] (u) -- (w);
  \draw[leg] (w) -- (x);
  \draw[leg] (x) -- (z);
  \draw[leg] (z) -- (v);

  \draw[line width=1.1pt,draw=hsand,dotted] (w) to[bend left=18] (z);
  \node[font=\footnotesize,text=hsand] at (5.6,3.05) {$H(w,z)$};

  \foreach \p in {u,w,x,z,v} \fill (\p) circle (2.2pt);
  \node[below=3.5pt] at (u) {$u$};
  \node[below=3.5pt] at (x) {$x$};
  \node[below=3.5pt] at (v) {$v$};
  \node[left=4.5pt] at (w) {$w$};
  \node[right=4.5pt] at (z) {$z$};

\end{tikzpicture}

    \caption{The estimate of \Cref{alg:k-alg}, where $w,z\in B(x)$,
    $w\in B(u)$, and $z\in B(v)$.}
    \label{fig:alg-query}
\end{figure}

First, we consider \emph{unweighted} graphs.
All previous algorithms~\cite{DBLP:conf/soda/ChechikHL25,kadria2026improved} operate as follows.
During preprocessing, they construct a table $H$, where
$H(w,z)$ is the minimum of $d(w,x)+d(x,z)$ over all vertices $x$
with $w,z\in B(x)$.
A query then examines every $(w,z)\in B(u)\times B(v)$ and returns as $\hat d(u,v)$ the minimum of
$d(u,w)+H(w,z)+d(z,v)$.
In particular, if for some vertex $x$ on $P(u,v)$ there are
$w\in B(u)\cap B(x)$ and $z\in B(v)\cap B(x)$, then these two bunch
elements provide a connection through $x$. We call $x$ a
\emph{connecting vertex} for $(u,v)$ (see \Cref{fig:alg-query}).
Constructing $H$ takes $\Ot(n^{1+2/k})$ additional time, and querying each pair takes $\Ot(n^{2/k})$ time, yielding an $\Ot(mn^{1/k} + n^{1+2/k})$-time algorithm.

Previous papers~\cite{DBLP:conf/soda/ChechikHL25,kadria2026improved} showed that this algorithm gives a $(\frac43 k,O(k))$-approximation, using the \emph{middle vertex} of $P(u,v)$, i.e., the vertex at equal distance from $u$ and $v$, as the connecting vertex.
We re-derive the $\frac 43k$ bound in a form that exposes its bottleneck, and then show how our analysis removes it.

Recall that $\nabla_i(v)=h_i(v)-h_{i-2}(v)$ is the two-level growth at $v \in V$.
Consider a pair $(u,v)$ with $d(u,v)=D$, where $D$ is even, and let $x$ be the middle vertex of $P(u,v)$, i.e., $d(u,x)=d(x,v)=D/2$.

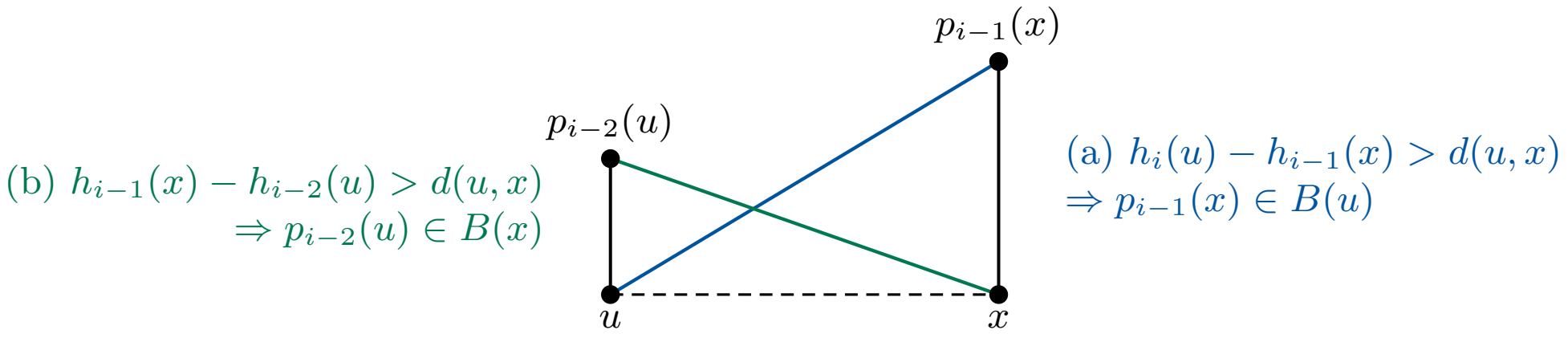
\begin{figure}[H]
    \centering
    \definecolor{gapa}{RGB}{0,84,159}
    \definecolor{gapb}{RGB}{0,120,80}
\begin{tikzpicture}[x=0.8cm,y=0.8cm,line width=0.75pt,
    every node/.style={font=\small,inner sep=1.5pt},
    casea/.style={line width=0.8pt,draw=gapa},
    caseb/.style={line width=0.8pt,draw=gapb}]

  \coordinate (u) at (0,0);
  \coordinate (pu) at (0,1.4);
  \coordinate (x) at (4,0);
  \coordinate (px) at (4,2.4);

  \draw[densely dashed,line width=0.6pt] (u) -- (x);

  \draw[casea] (u) -- (px);
  \draw (px) -- (x);

  \draw (u) -- (pu);
  \draw[caseb] (pu) -- (x);

  \foreach \p in {u,x,pu,px} \fill (\p) circle (2.2pt);
  \node[below=2pt] at (u) {$u$};
  \node[below=2pt] at (x) {$x$};
  \node[above=2pt] at (pu) {$p_{i-2}(u)$};
  \node[above=2pt] at (px) {$p_{i-1}(x)$};

  \node[anchor=east,align=right,text=gapb] at (-0.6,0.9)
      {(b) $h_{i-1}(x)-h_{i-2}(u)>d(u,x)$\\ $\Rightarrow p_{i-2}(u)\in B(x)$};
  \node[anchor=west,align=left,text=gapa] at (4.6,1.2)
      {(a) $h_i(u)-h_{i-1}(x)>d(u,x)$\\ $\Rightarrow p_{i-1}(x)\in B(u)$};
\end{tikzpicture}

    \caption{\Cref{cl:pivot-connection}. Since $\nabla_i(u)={\color{gapa}\bigl(h_i(u)-h_{i-1}(x)\bigr)}+{\color{gapb}\bigl(h_{i-1}(x)-h_{i-2}(u)\bigr)}>2d(u,x)$,
    at least one of the two terms is larger than $d(u,x)$, and hence one of
    the paths {\color{gapa}(a)} and {\color{gapb}(b)} exists.}
    \label{fig:two-level-gap}
\end{figure}

The middle-vertex analysis checks when $x$ is a ``good'' connecting vertex for the pair $(u,v)$. Assume that $\nabla_i(u)>2d(u,x)=D$ for some level $i$. Then
\[
\nabla_i(u)=\bigl(h_i(u)-h_{i-1}(x)\bigr)+\bigl(h_{i-1}(x)-h_{i-2}(u)\bigr)>2d(u,x),
\]
and therefore $h_i(u)-h_{i-1}(x)>d(u,x)$ or $h_{i-1}(x)-h_{i-2}(u)>d(u,x)$. By \Cref{lem:tz-bunch}, either $p_{i-1}(x)\in B(u)$ or $p_{i-2}(u)\in B(x)$ (see \Cref{fig:two-level-gap}).
Until such a growth occurs, the pivot distance grows by at most $D$ every two levels. The same argument applies independently if $\nabla_j(v)>2d(x,v)=D$. If such a growth occurs at both endpoints, possibly at different levels, then $x$ serves as a connecting vertex for $(u,v)$.
Combining these middle-vertex connections with the classical Thorup--Zwick search gives the $(\frac 43 k, O(k))$-approximation. Roughly, if the pivot distances grow by about $D/2$ per level up to some level $r$, a connection through $x$ at level $r$ gives an estimate of about $2rD$, while continuing the Thorup--Zwick search from level $r$ gives about $(2k-r)D$. Balancing the two at $r=\frac23k$ gives $\frac43k$.
If we consider only the middle vertex as a possible connecting vertex, then this analysis is tight.

The middle-vertex analysis has an inherent limitation: since $x$ is fixed in advance, it requires sufficient growth at \emph{each} endpoint separately.
To see this, note that the argument above shows that a two-level growth $\nabla_i(u)$ at $u$ lets $u$ connect to every vertex $y$ of $P(u,v)$ with $d(u,y)<\nabla_i(u)/2$. We say that $u$ \emph{reaches} $y$. Similarly, $v$ reaches every vertex $y$ with $d(y,v)<\nabla_j(v)/2$, and a vertex reached by both endpoints is a connecting vertex.
Since the middle vertex satisfies $d(u,x)=d(x,v)=D/2$, the previous analyses must wait for a level $i$ with $\nabla_i(u)>D$ \emph{and} a level $j$ with $\nabla_j(v)>D$. Extra growth at one endpoint cannot compensate for missing growth at the other.

\begin{figure}[h]
    \centering
    \tikzset{every picture/.style={line width=0.75pt}} 
%

\begin{tikzpicture}[x=0.62cm,y=0.62cm,line width=0.75pt,
    every node/.style={font=\small,inner sep=1.5pt},
    dim/.style={|<->|,>=stealth,line width=0.6pt},
    drop/.style={densely dotted,line width=0.5pt,gray}]

  \fill[gray!20] (6.5,-1.7) rectangle (8,1.9);

  \draw (0,0) -- (10,0);
  \foreach \t in {1,...,9} \fill (\t,0) circle (1.3pt);
  \fill (0,0) circle (2.2pt);
  \fill (10,0) circle (2.2pt);
  \fill (7,0) circle (2.2pt);
  \fill (5,0) circle (2.2pt);

  \node[below=2pt] at (0,0) {$u$};
  \node[below right=1pt] at (10,0) {$v$};
  \node[above=2pt] at (7,0) {$y$};
  \node[below=2pt] at (5,0) {$x$};

  \draw[drop] (0,0) -- (0,2.2);
  \draw[drop] (8,0) -- (8,2.2);
  \draw[dim] (0,2.2) -- (8,2.2)
      node[midway,above=1pt] {$\nabla_i(u)/2$};
  \node[above=1pt,font=\footnotesize] at (4,3.3) {$F>0$};

  \draw[drop] (10,0) -- (10,-2.0);
  \draw[drop] (6.5,0) -- (6.5,-2.0);
  \draw[dim] (6.5,-2.0) -- (10,-2.0)
      node[midway,below=1pt] {$\nabla_j(v)/2$};
  \node[below=1pt,font=\footnotesize] at (8.25,-3.1) {$G>0$};

  \node[font=\footnotesize,align=center,anchor=north] at (2.4,-1.4)
      {$\nabla_i(u)+\nabla_j(v)\ge 2D+3$\\ forces the two ranges to overlap};
\end{tikzpicture}

    \caption{The Mixed-Level Growth Lemma (\Cref{lem:mixed-level-growth}). The ranges $F>0$ and $G>0$ overlap (shaded). The middle vertex $x$ is outside the overlap, while every vertex $y$ in the overlap satisfies the conditions of \Cref{cl:connecting-vertex-exists}.}
    \label{fig:mixed-level-growth}
\end{figure}
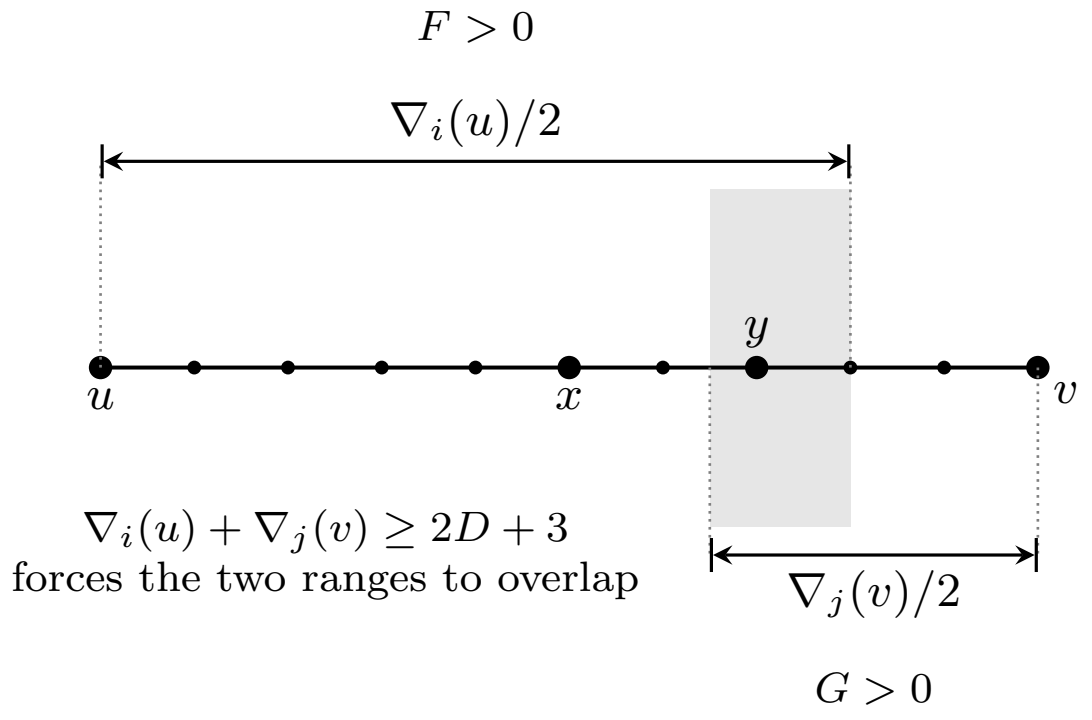

For example, suppose that $\nabla_i(u)=2D-1$ and $\nabla_j(v)=D-1$.
Then $u$ reaches every vertex at distance at most $D-1$ from it, which is almost all of $P(u,v)$, while $v$ reaches only the vertices at distance at most $D/2-1$ from it, which excludes the middle vertex.
Hence, the middle vertex is not a connecting vertex for the levels $i$ and $j$.
Nevertheless, the portion of $P(u,v)$ reached from $u$ and the portion of $P(u,v)$ reached from $v$ overlap, and every vertex in the overlap is a connecting vertex (see \Cref{fig:mixed-level-growth}).

This is the main new idea behind the Mixed-Level Growth Lemma. We do not fix the connecting vertex in advance. Instead, we consider all vertices of $P(u,v)$ and choose the connecting vertex according to the growth at both endpoints.
Such a vertex exists whenever the two reached portions together cover $P(u,v)$, which happens if $\nabla_i(u)+\nabla_j(v)\ge 2D+3$.
Thus, only the \emph{total} growth at the two endpoints matters, and not how it is split between them.
This seemingly small change affects only the analysis, not the algorithm. Nevertheless, it suffices to obtain the $(k,k-1)$-approximation of \Cref{thm:intro-unweighted-npsp}.
We formalize this in the Mixed-Level Growth Lemma below.

\paragraph{Tool I: The Mixed-Level Growth Lemma.}
In an unweighted graph, if for some $2\le i,j<k$,
\begin{equation}
\label{eq:overview-mixed-level-growth}
\nabla_i(u)+\nabla_j(v)\geq 2D+3,
\end{equation}
then some vertex $x$ is a connecting vertex for $(u,v)$, and the estimate $\hat d(u,v)$ returned by the query satisfies
\begin{equation}
\label{eq:overview-mixed-level-estimate}
\hat d(u,v)\leq 2h_{i-2}(u)+2h_{j-2}(v)+3D.
\end{equation}

The growth at each endpoint determines how far the connecting vertex
can be from it. Instead of fixing $x$ at the middle vertex, we let $x$ be the last vertex of $P(u,v)$ with $d(u,x)<\nabla_i(u)/2$, and show that if Condition~\eqref{eq:overview-mixed-level-growth} is met, then $\nabla_i(u)>2d(u,x)$ and $\nabla_j(v)>2d(x,v)$. Thus, we can use the same proof idea as for the middle vertex to show that this $x$ is a good connecting vertex for $(u,v)$.

The lemma, together with the last-level Thorup--Zwick estimate
$\hat d(u,v)\le 2\min\{h_{k-1}(u),h_{k-1}(v)\}+D$, reduces the remaining analysis to a numerical problem.
For any pair of levels $i,j$ at which
\eqref{eq:overview-mixed-level-estimate} would give a sufficiently short
estimate, the absence of such an estimate forces
$\nabla_i(u)+\nabla_j(v)<2D+3$.
Combining these growth constraints with the last-level
estimate yields the bound $kD+k-1$, while preserving the previous running time.

\paragraph{Removing the extra running-time term.}
The extra running time arises in two places. First, the preprocessing stores $H(w,z)$ for all pairs $w,z\in B(x)$, for every vertex $x$, which takes $\Ot(n^{1+2/k})$ time. Second, the query examines all pairs in $B(u)\times B(v)$, which takes $\Ot(n^{2/k})$ time per query, and $\Ot(n^{1+2/k})$ time overall. We change both parts.

\paragraph{The query.} Instead of all of $B(u)\times B(v)$, the query examines only the \emph{mixed} pairs
\begin{equation}
\label{eq:overview-mixed-queries}
\{p_i(u) \mid 0\le i<k\}\times B(v)
\qquad\text{and}\qquad
B(u)\times \{p_j(v) \mid 0\le j<k\},
\end{equation}
which takes $\Ot(n^{1/k})$ time per pair.

\paragraph{The preprocessing.} Pairs of bunch elements of a common vertex are no longer stored. Instead, we use the preprocessing technique of~\cite{kadria2026improved}, which also handles the difficulty that weighted graphs introduce: the ideal position of the connecting vertex may lie inside a heavy edge, and moving it to either endpoint may exhaust the available growth.
For every oriented edge $(x,y)$, every level $i$, and every $z\in B(y)$,
we update $H(p_i(x),z)$ and $H(z,p_i(x))$ with the length of the walk
$p_i(x)\to x\to y\to z$, which crosses the edge $(x,y)$ directly. This takes $\Ot(mn^{1/k})$ time.

\begin{figure}[t]
    \centering
    \definecolor{bunchu}{RGB}{0,84,159}
    \definecolor{bunchx}{RGB}{0,120,80}
    \definecolor{bunchv}{RGB}{176,32,32}
    \definecolor{bunchy}{RGB}{112,48,160}
    \definecolor{hsand}{RGB}{184,145,82}
    \tikzset{every picture/.style={line width=0.75pt}}
\tikzset{
  lbpic/.style={x=0.64cm,y=0.8cm,line width=0.75pt,
    every node/.style={font=\small,inner sep=1.5pt}},
  leg/.style={line width=0.7pt},
  entry/.style={line width=1.1pt,draw=hsand,dotted},
  bunchlabel/.style={font=\footnotesize},
  bunchpad/.style={line width=0.75cm,line join=round,line cap=round}}
\begin{minipage}[t]{0.49\textwidth}
\centering
\begin{tikzpicture}[lbpic]
  \coordinate (u) at (0,0);
  \coordinate (pu) at (0,2.6);
  \coordinate (x) at (3.4,0);
  \coordinate (y) at (5.0,0);
  \coordinate (py) at (5.0,2.6);
  \coordinate (v) at (8.4,0);

  \begin{scope}[transparency group,opacity=0.13]
    \filldraw[bunchx,bunchpad] (pu) -- (py) -- (y) -- (x) -- cycle;
  \end{scope}
  \begin{scope}[transparency group,opacity=0.15]
    \filldraw[bunchv,bunchpad] (py) -- (v |- py) -- (v) -- cycle;
  \end{scope}
  \node[text=bunchx,bunchlabel,anchor=south west] at (-0.55,3.05) {$B(x)\cup\{p_j(y)\mid j\in[k]\}$};
  \node[text=bunchv,bunchlabel,anchor=south east] at (8.95,3.05) {$B(v)$};

  \draw[line width=0.8pt,densely dashed] (u) -- (x) (y) -- (v);

  \draw[leg] (u) -- (pu);
  \draw[leg] (pu) -- (x);
  \draw[leg] (x) -- (y);
  \draw[leg] (y) -- (py);
  \draw[leg] (py) -- (v);

  \draw[entry] (pu) to[bend left=18] (py);
  \node[font=\footnotesize,anchor=south,text=hsand] at (2.5,2.0) {$H(p_i(u),p_j(y))$};

  \foreach \p in {u,pu,x,y,py,v} \fill (\p) circle (2.2pt);
  \node[below=3.5pt] at (u) {$u$};
  \node[below=3.5pt] at (x) {$x$};
  \node[below=3.5pt] at (y) {$y$};
  \node[below=3.5pt] at (v) {$v$};
  \node[left=4.5pt] at (pu) {$p_i(u)$};
  \node[above right=1.5pt] at (py) {$p_j(y)$};

\end{tikzpicture}

\small (a)
\end{minipage}\hfill
\begin{minipage}[t]{0.49\textwidth}
\centering
\begin{tikzpicture}[lbpic]
  \coordinate (u) at (0,0);
  \coordinate (px) at (3.4,2.6);
  \coordinate (x) at (3.4,0);
  \coordinate (y) at (5.0,0);
  \coordinate (pv) at (8.4,2.6);
  \coordinate (v) at (8.4,0);

  \begin{scope}[transparency group,opacity=0.15]
    \filldraw[bunchu,bunchpad] (u |- px) -- (px) -- (u) -- cycle;
  \end{scope}
  \begin{scope}[transparency group,opacity=0.13]
    \filldraw[bunchy,bunchpad] (px) -- (pv) -- (y) -- (x) -- cycle;
  \end{scope}
  \node[text=bunchu,bunchlabel,anchor=south west] at (-0.55,3.05) {$B(u)$};
  \node[text=bunchy,bunchlabel,anchor=south east] at (8.95,3.05) {$B(y)\cup\{p_i(x)\mid i\in[k]\}$};

  \draw[line width=0.8pt,densely dashed] (u) -- (x) (y) -- (v);

  \draw[leg] (u) -- (px);
  \draw[leg] (px) -- (x);
  \draw[leg] (x) -- (y);
  \draw[leg] (y) -- (pv);
  \draw[leg] (pv) -- (v);

  \draw[entry] (px) to[bend left=18] (pv);
  \node[font=\footnotesize,anchor=south,text=hsand] at (5.9,2.0) {$H(p_i(x),p_j(v))$};

  \foreach \p in {u,x,px,y,v,pv} \fill (\p) circle (2.2pt);
  \node[below=3.5pt] at (u) {$u$};
  \node[below=3.5pt] at (x) {$x$};
  \node[below=3.5pt] at (y) {$y$};
  \node[below=3.5pt] at (v) {$v$};
  \node[above left=1.5pt] at (px) {$p_i(x)$};
  \node[right=4.5pt] at (pv) {$p_j(v)$};

\end{tikzpicture}

\small (b)
\end{minipage}

    \caption{The two walk shapes used in the analysis of \Cref{alg:65k-alg}, where
    $(x,y)\in P(u,v)$. In (a), $p_i(u)\in B(x)$ and $p_j(y)\in B(v)$. In (b),
    $p_i(x)\in B(u)$ and $p_j(v)\in B(y)$. The dotted edge is the entry of $H$ set for $(x,y)$.}
    \label{fig:lb-estimate}
\end{figure}
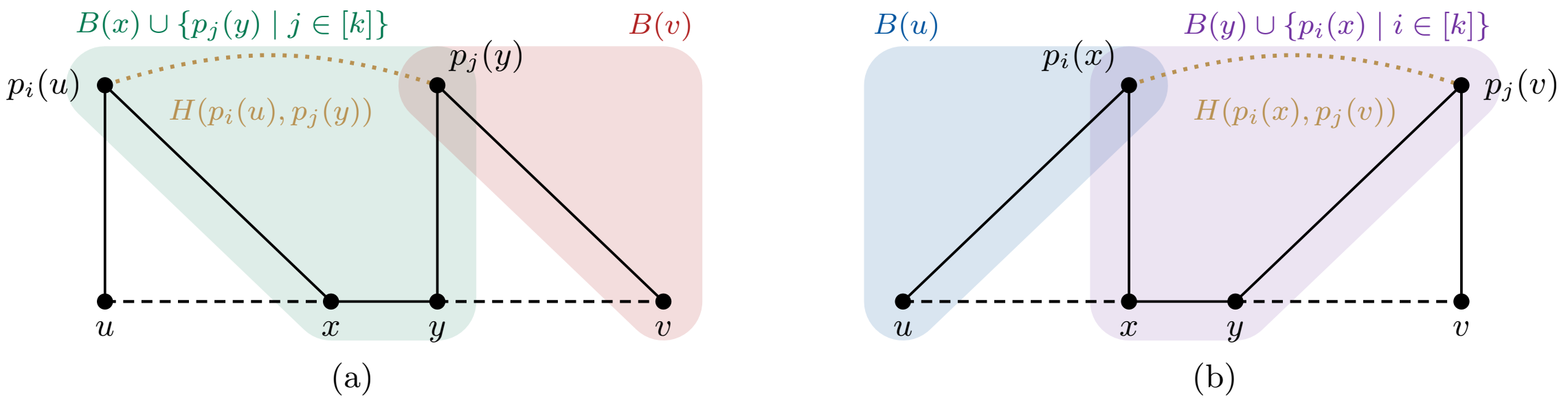

Every estimate that the query returns is the length of a $u$--$v$ walk, and our analysis uses only walks of the forms
\[
u\to p_i(u)\to x\to y\to p_j(y)\to v
\qquad\text{or}\qquad
u\to p_i(x)\to x\to y\to p_j(v)\to v,
\]
where $(x,y)$ is an edge of $G$ (see \Cref{fig:lb-estimate}). The total running time is $\Ot(mn^{1/k})$ for $m$ pairs.

The main challenge is to prove that the mixed connections alone
suffice for a small stretch. For this, we develop the following two tools.

\paragraph{Tool II: The Two-Level Growth Lemma.}
If for some $2\le i<k$,
\begin{equation}
\label{eq:overview-two-level-growth}
h_i(u)-h_{i-2}(v)>D,
\end{equation}
then
\begin{equation}
\label{eq:overview-two-level-estimate}
\hat{d}(u,v)\le 2h_{i-2}(v)+h_{i-1}(u)+h_i(u)+D.
\end{equation}
The same statement holds with $u$ and $v$ exchanged.

The lemma splits a gap across the two endpoints rather than at a single endpoint. For every $z\in P(u,v)$,
\[
h_i(u)-h_{i-2}(v)=\bigl(h_i(u)-h_{i-1}(z)\bigr)+\bigl(h_{i-1}(z)-h_{i-2}(v)\bigr)>d(u,z)+d(z,v),
\]
so by \Cref{lem:tz-bunch}, $p_{i-1}(z)\in B(u)$ or $p_{i-2}(v)\in B(z)$.
Moving along $P(u,v)$, we find an edge $(x,y)$ with $p_{i-1}(x)\in B(u)$ and $p_{i-2}(v)\in B(y)$.
Unlike in the Mixed-Level Growth Lemma, the pivot used on each side is fixed in advance, and the two always form a mixed pair. This gives the estimate of type~(b) in \Cref{fig:lb-estimate}.

The lemma helps us control the growth of the lower levels.
Whenever \eqref{eq:overview-two-level-estimate} would give a sufficiently short estimate, the absence
of such an estimate forces Condition~\eqref{eq:overview-two-level-growth} to fail in both orientations: \( h_i(u)\le h_{i-2}(v)+D \) and \( h_i(v)\le h_{i-2}(u)+D\). Adding these inequalities gives
\[
\nabla_i(u)+\nabla_i(v)\le 2D,
\]
which is the constraint of the Mixed-Level Growth Lemma for $j=i$, now obtained from mixed pairs alone and in weighted graphs.
Thus, $u$ and $v$ together grow by at most $2D$ every two levels, that is, by about $D/2$ per level per endpoint on average.
For pivot distances around $iD/2$, each endpoint contributes about $iD$ to \eqref{eq:overview-two-level-estimate}, since the lemma pairs level $i$ at $u$ with level $i-2$ at $v$.
The total is about $2iD$, which is below the target of about $\frac 65kD$ only for $i\lesssim\frac 35k$.
To handle the higher levels, we need a lemma that pairs a high level at one endpoint with a low level at the other, so that only one endpoint contributes a large term.
The Long--Short Growth Lemma does exactly this. It combines a two-level growth at level $i$ of one endpoint with a one-level growth at an arbitrary level $j$ of the other, and gives an estimate of about $(i+j)D$.

\paragraph{Tool III: The Long--Short Growth Lemma.}
Recall that $\Delta_i(u)=h_i(u)-h_{i-1}(u)$ is the one-level growth at $u$, and that $\nabla_i(u)=\Delta_i(u)+\Delta_{i-1}(u)$ is the two-level growth at $u$. 
The Long--Short Growth Lemma states that if
\begin{equation}
\label{eq:overview-long-short-growth}
\nabla_i(u)+\Delta_j(v)>2D,
\end{equation}
then
\begin{equation}
\label{eq:overview-long-short-estimate}
\hat d(u,v)\le 2h_{i-1}(u)+2h_{j-1}(v)+3D.
\end{equation}
The same statement holds with $u$ and $v$ exchanged.

\begin{figure}[H]
    \centering
    \tikzset{every picture/.style={line width=0.75pt}}
%
\begin{tikzpicture}[x=0.9cm,y=0.9cm,line width=0.75pt,
    every node/.style={font=\small,inner sep=1.5pt},
    dim/.style={|<->|,>=stealth,line width=0.6pt},
    drop/.style={densely dotted,line width=0.5pt,gray}]

  \fill[gray!20] (6.0,-1.3) rectangle (7.0,1.4);

  \draw (0,0) -- (10,0);
  \foreach \t in {1.2,2.4,3.4,4.5,9.3} \fill (\t,0) circle (1.3pt);
  \foreach \t in {0,5.5,8.5,10} \fill (\t,0) circle (2.2pt);
  \draw[line width=1.6pt,black] (6.5,0) +(-2.8pt,-2.8pt) -- +(2.8pt,2.8pt) +(-2.8pt,2.8pt) -- +(2.8pt,-2.8pt);

  \node[below=2pt] at (0,0) {$u$};
  \node[below=2pt] at (5.5,0) {$x$};
  \node[below=2pt] at (8.5,0) {$y$};
  \node[below=2pt] at (10,0) {$v$};
  \node[above=3.5pt] at (6.5,0) {$z$};

  \draw[drop] (0,0) -- (0,1.7);
  \draw[drop] (7.0,0) -- (7.0,1.7);
  \draw[dim] (0,1.7) -- (7.0,1.7)
      node[midway,above=1pt] {$\nabla_i(u)/2$};

  \draw[drop] (10,0) -- (10,-1.6);
  \draw[drop] (6.0,0) -- (6.0,-1.6);
  \draw[dim] (6.0,-1.6) -- (10,-1.6)
      node[midway,below=1pt] {$\Delta_j(v)/2$};
\end{tikzpicture}

    \caption{\Cref{cl:long-short-split-edge}. The two ranges overlap at a point
    $z$ of the edge $(x,y)$, which may be heavy. The point $z$ is not a
    vertex but a point inside $(x,y)$ that is used only in the analysis.}
    \label{fig:long-short}
\end{figure}
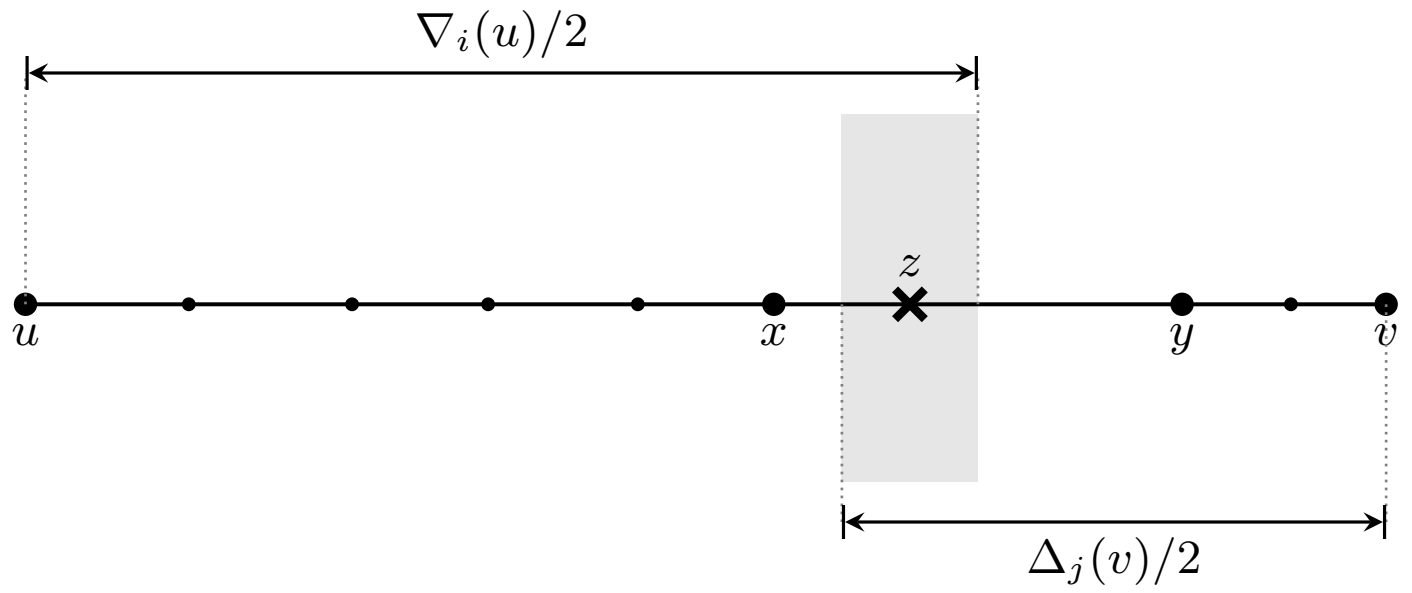

The intuition is to first handle a long part of $P(u,v)$,
and then use the short remaining part to complete a mixed estimate.
Condition~\eqref{eq:overview-long-short-growth} says that the range of the long gap,
$d(u,\cdot)<\nabla_i(u)/2$, and the range of the short gap,
$d(\cdot,v)<\Delta_j(v)/2$, together cover $P(u,v)$, so they meet at
some point $z$ (see \Cref{fig:long-short}).
In an unweighted graph, we may take $z$ to be a vertex and set $x=y=z$, at the cost of increasing the threshold $2D$ in Condition~\eqref{eq:overview-long-short-growth} by an additive constant, as in Tool~I.
In a weighted graph, $z$ may lie inside a heavy edge $(x,y)$. In this case, we let
the long part end at $x$ and the short part start at $y$, so that
$\nabla_i(u)>2d(u,x)$ and $\Delta_j(v)>2d(y,v)$, and use the entry of
$H$ recorded for $(x,y)$ to cross the edge at no extra cost.

On the long $u$--$x$ part, as before, $\nabla_i(u)>2d(u,x)$ guarantees
either a pivot of $u$ in $B(x)$ or a pivot of $x$ in $B(u)$.
We allow either outcome, and then look on the short $y$--$v$ part
for the pivot that completes it: a pivot of $y$ in $B(v)$ in the first
case, and a pivot of $v$ in $B(y)$ in the second.

To be ready for both outcomes, the short part uses the one-level
condition $\Delta_j(v)>2d(y,v)$, which guarantees both pivots at once
(see \Cref{lem:one-level-gap}).
This is more expensive, as the growth must now occur within a single
level rather than over two levels, but since the $y$--$v$ part is
short, the extra cost is small.
The two outcomes give exactly the two walk shapes of \Cref{fig:lb-estimate}.

\paragraph{Obtaining the $\frac 65 k$ bound.}
The two lemmas constrain different parts of the growth.
The Two-Level Growth Lemma limits growth at lower levels, where the
sum of the pivot distances is still small.
The Long--Short Growth Lemma then relates later two-level increments
to earlier one-level increments.
Each constraint applies whenever the corresponding estimate would
be sufficiently short.
Combining these constraints with the last-level Thorup--Zwick estimate
reduces the analysis to a numerical optimization problem, whose solution
guarantees an estimate of at most $(\frac 65 k+7)D$,
giving the claimed stretch in $\Ot(mn^{1/k})$ total time.
In \Cref{sec:65k-optimization}, we give the intuition behind this solution and show that the constant $\frac 65$ is tight for the estimates of our algorithm. See also \Cref{fig:growth-profile} for a hard instance.

\paragraph{Adapting to other problems.} We presented our tools through \nPSP and \mPSP, where their application is
simplest to explain. The lemmas themselves describe general properties
of the Thorup--Zwick hierarchy and are not specific to these problems.

Our modification for distance oracles in weighted graphs is simple: the preprocessing is independent of the queried pairs, and therefore we compute and store $H$ in the preprocessing phase and compute the estimate upon receiving a query. For unweighted graphs, we show that one can reduce the space to $\Ot(n^{1+1/k})$ by not storing the edge-related distances, at the cost of an additional $O(k)$ additive error.

To cover $ST$-SP and MSSP, we study the more general $(t,s)$-PSP problem, in which each of the $t$ requested pairs has an endpoint in a set $A$ of $s$ sources. Following the parameterized distance-oracle construction of~\cite{kadria2025fasteralgorithms2k1stretchdistance,RodittyTZ05}, we set $A_0=A$ and sample the remaining levels as usual. Because our growth lemmas depend only on pivot-distance increments at the queried endpoints and, in weighted graphs, on differences between the pivot distances of the two endpoints, the analysis carries over with at most $2d(u,v)$ additional error in unweighted graphs and at most $4d(u,v)$ in weighted graphs. This yields running times $\Ot((m+t)s^{1/k})$ for weighted graphs and $\Ot(ms^{1/k}+(n+t)s^{2/k})$ for unweighted graphs.

\section{$(k,k-1)$-approximation in $\widetilde{O}(mn^{1/k}+n^{1+2/k})$ time}
\label{sec:k-stretch}

In this section, we provide a $(k,k-1)$-approximation algorithm for \nPSP that runs in $\Ot(mn^{1/k}+n^{1+2/k})$ time. The algorithm is the same as that of~\cite{kadria2026improved}, which is a simplification of~\cite{DBLP:conf/soda/ChechikHL25}. The main contribution of this section is a more precise analysis of that algorithm, which establishes the $(k,k-1)$-approximation guarantee.

\Reminder{thm:intro-unweighted-npsp}

Throughout this section, let $I$ be the set of $n$ queried vertex pairs.

\subsection{The previous (and current) algorithm}

The algorithm works as follows. First, the algorithm initializes two empty hash tables $H$ and $\hat{d}$, and for every $u\in V$ computes $B(u)$ using \Cref{lem:tz-construction}.\footnote{$H$ returns $\infty$ on unset pairs, and $H(w,w)=0$ for every $w$.} Then, for every $x\in V$ and every $w,z\in B(x)$, the algorithm sets $H(w,z)$ to $\min(H(w,z), d(x,w)+d(x,z))$. Next, for every vertex pair $\pair{u,v}\in I$ and every $w\in B(u)$ and $z\in B(v)$, the algorithm sets $\hat{d}(u,v)$ to $\min(\hat{d}(u,v), d(u,w)+H(w,z)+d(z,v))$. The algorithm returns $\hat{d}(u,v)$ for every $\pair{u,v}\in I$. Pseudocode for the algorithm is presented in \Cref{alg:k-alg}, and the estimate that it evaluates is illustrated in \Cref{fig:alg-query}.
\begin{algorithm2e}
    \caption{$(k,k-1)$-approximation for \nPSP}\label{alg:k-alg}
    $H \gets \HashTable(); \hat{d} \gets \HashTable()$ \\
    Compute $B(u)$ for every $u\in V$ [\Cref{lem:tz-construction}]\\
    \For{$x\in V$} {
        \For{$w,z\in B(x)$} {
            $H(w,z) = \min(H(w,z), d(x,w)+d(x,z))$
        }
    }
    \For{$\pair{u,v}\in I$} {
        \For{$w\in B(u)$ and $z\in B(v)$} {
            $\hat{d}(u,v) = \min(\hat{d}(u,v), d(u,w)+H(w,z)+d(z,v))$
        }
    } 
    \Return $\{\hat{d}(u,v) \mid \pair{u,v}\in I\}$
\end{algorithm2e}

\subsection{Our analysis}

For the rest of this section, fix a pair $\pair{u,v}\in I$, and let $D=d(u,v)$. Recall the two-level growth $\nabla_i(x)=h_i(x)-h_{i-2}(x)$ from \Cref{sec:prelim}.
Our analysis avoids fixing the connecting vertex on $P(u,v)$ in advance. Given levels $i$ and $j$, we choose a vertex $x\in P(u,v)$ according to the relative values of $\nabla_i(u)$ and $\nabla_j(v)$. The following lemma shows that it is enough for their sum to be large: the single condition $\nabla_i(u)+\nabla_j(v)\ge 2D+3$ guarantees a short estimate. Thus, a larger increase at one endpoint can compensate for a smaller increase at the other endpoint.

\begin{lemma}[Mixed-Level Growth Lemma]\label{lem:mixed-level-growth}
    Let $2\le i,j<k$, and let $D=d(u,v)$. If
    \begin{equation}\label{eq:mixed-level-growth-condition}
        \nabla_i(u)+\nabla_j(v)\ge 2D+3,
    \end{equation}
    then
    \(
        \hat{d}(u,v)\le 2h_{i-2}(u)+2h_{j-2}(v)+3D.
    \)
\end{lemma}

\begin{proof}

We divide the proof into three claims. First, we show that a single large two-level growth already gives the desired approximation.

\begin{cclaim}
If for some $2\le i<k$, $\nabla_i(u)>2D$, then $\hat d(u,v)\le 2h_{i-2}(u)+3D$. This holds symmetrically for $v$.
\end{cclaim}

\begin{proof}
Suppose first that $h_{i-1}(v)>D+h_{i-2}(u)$. By \Cref{lem:tz-bunch}, $p_{i-2}(u)\in B(v)$. Moreover, the triangle inequality gives
$h_{i-1}(u)\ge h_{i-1}(v)-D>h_{i-2}(u)$, so $p_{i-2}(u)\in B(u)$ as well. Therefore,
\[
\hat d(u,v)
\le d(u,p_{i-2}(u))+d(p_{i-2}(u),v)
\le 2h_{i-2}(u)+D.
\]

Otherwise, $h_{i-1}(v)\le D+h_{i-2}(u)$. Since $\nabla_i(u)>2D$, we have
\[
h_i(u)>h_{i-2}(u)+2D\ge h_{i-1}(v)+D.
\]
By \Cref{lem:tz-bunch}, $p_{i-1}(v)\in B(u)$. Also,
$h_i(v)\ge h_i(u)-D>h_{i-1}(v)$, so $p_{i-1}(v)\in B(v)$. Hence,
\[
\hat d(u,v)
\le d(u,p_{i-1}(v))+d(p_{i-1}(v),v)
\le D+2h_{i-1}(v)
\le 2h_{i-2}(u)+3D.
\]
\end{proof}

By the previous claim, we may assume that
$\nabla_i(u),\nabla_j(v)\le 2D$. Together with \eqref{eq:mixed-level-growth-condition}, this implies that $\nabla_i(u),\nabla_j(v)>0$.

\begin{cclaim}\label{cl:connecting-vertex-exists}
There exists a vertex $x\in P(u,v)$ such that
\begin{align*}
\nabla_i(u)>2d(u,x), \quad \text{and} \quad
\nabla_j(v)>2d(v,x).
\end{align*}
\end{cclaim}
\begin{proof}
For every vertex $q\in P(u,v)$, we define
\begin{align*}
F(q)=\nabla_i(u)-2d(u,q), \quad \text{and} \quad
G(q)=\nabla_j(v)-2d(v,q).
\end{align*}
Our goal is to find a vertex $x$ such that $F(x)>0$ and $G(x)>0$. See \Cref{fig:mixed-level-growth} for an illustration.

Let $x$ be the last vertex of $P(u,v)$, starting from $u$, such that $F(x)>0$. Such a vertex exists since $\nabla_i(u)>0$, so $F(u)>0$. Moreover, since $\nabla_i(u)\le 2D$, we have $F(v)=\nabla_i(u)-2D\le 0$, and therefore $x$ comes before $v$ on the path.

Since the graph is unweighted, $F$ decreases by $2$ at each edge traversed from $u$ to $v$. Since $x$ is the last vertex such that $F(x)>0$, we have $0<F(x)\le 2$. By \eqref{eq:mixed-level-growth-condition},
\begin{align*}
2+G(x) \ge F(x)+G(x)=\nabla_i(u)+\nabla_j(v)-2D \ge 3.
\end{align*}
Thus, $G(x)\ge1$, and $x$ satisfies both required inequalities.
\end{proof}

We show that $x$ can act as a connecting vertex. For this, we choose between $p_{i-2}(u)$ and $p_{i-1}(x)$, using the former whenever
$d(u,x)+h_{i-2}(u)<h_{i-1}(x)$ (see \Cref{fig:two-level-gap}).

\begin{cclaim}\label{cl:pivot-connection}
There exists $w_u\in\{p_{i-2}(u),p_{i-1}(x)\}$ such that $w_u\in B(u)\cap B(x)$ and
\[
d(u,w_u)+d(w_u,x)\le 2h_{i-2}(u)+3d(u,x).
\]
\end{cclaim}
\begin{proof}
Suppose first that $d(u,x)+h_{i-2}(u)<h_{i-1}(x)$, and choose $w_u=p_{i-2}(u)$. By \Cref{lem:tz-bunch}, $w_u\in B(x)$. Moreover,
\[
h_{i-1}(u)\ge h_{i-1}(x)-d(u,x)>h_{i-2}(u),
\]
so $w_u\in B(u)$ as well. The triangle inequality gives
\[
d(u,w_u)+d(w_u,x)\le 2h_{i-2}(u)+d(u,x).
\]

Otherwise, $h_{i-1}(x)\le h_{i-2}(u)+d(u,x)$, and we choose $w_u=p_{i-1}(x)$. By \Cref{cl:connecting-vertex-exists},
\[
h_i(u)>h_{i-2}(u)+2d(u,x)\ge h_{i-1}(x)+d(u,x).
\]
Thus, \Cref{lem:tz-bunch} gives $w_u\in B(u)$. Also,
$h_i(x)\ge h_i(u)-d(u,x)>h_{i-1}(x)$, so $w_u\in B(x)$. Finally,
\[
d(u,w_u)+d(w_u,x)
\le d(u,x)+2h_{i-1}(x)
\le 2h_{i-2}(u)+3d(u,x),
\]
as required.
\end{proof}

Choose $w_u$ as in \Cref{cl:pivot-connection}, and symmetrically choose $w_v\in B(v)\cap B(x)$ such that
\[
d(v,w_v)+d(w_v,x)\le 2h_{j-2}(v)+3d(v,x).
\]

Since $w_u,w_v\in B(x)$, the algorithm ensures that
$H(w_u,w_v)\le d(w_u,x)+d(x,w_v)$.
Moreover, $w_u\in B(u)$ and $w_v\in B(v)$, so the query considers this pair and gives
\begin{align*}
\hat d(u,v)
&\le d(u,w_u)+H(w_u,w_v)+d(w_v,v)\\
&\le d(u,w_u)+d(w_u,x)+d(x,w_v)+d(w_v,v)\\
&\le 2h_{i-2}(u)+3d(u,x)+2h_{j-2}(v)+3d(v,x)\\
&=2h_{i-2}(u)+2h_{j-2}(v)+3D,
\end{align*}
as required.
\end{proof}

Next, we use \Cref{lem:mixed-level-growth} to obtain a good approximation ratio for the algorithm.
We have the following two bounds.
\begin{enumerate}
    \item $\hat d(u,v)\le 2\min(h_{k-1}(u),h_{k-1}(v))+d(u,v)$.
    \item If $\nabla_i(u)+\nabla_j(v)\ge 2d(u,v)+3$, then $\hat d(u,v)\le 2h_{i-2}(u)+2h_{j-2}(v)+3d(u,v)$.
\end{enumerate}

At this point, all graph-theoretic aspects of the analysis are captured by the two bounds above, and it remains only to solve a purely numerical optimization problem. We first assume that $k$ is odd, and set $x_i=\nabla_{2i}(u)$, $y_i=\nabla_{2i}(v)$, and $\kappa=(k-1)/2$.

\begin{lemma}\label{lem:k-optimization}\Copy{lem:k-optimization}{
Let $D\ge0$ and $\kappa\ge0$ be integers, and let $c\ge0$ be a real number. Let $x_i,y_i$ be nonnegative integers for every $1\le i\le\kappa$. Then
\[
\min\left(
\sum_{i=1}^{\kappa}x_i,
\sum_{i=1}^{\kappa}y_i+2c,
\min_{\substack{1\le i,j\le\kappa\\
x_i+y_j\ge 2D+3}}
\left\{
\sum_{r<i}x_r+\sum_{r<j}y_r+D
\right\}
\right)
\le \kappa D+\kappa+c.
\]
}
\end{lemma}

We defer the proof of the lemma to \Cref{app:numerical-lemma}.

In our algorithm, $\hat d(u,v)$ is at most twice the optimization target of \Cref{lem:k-optimization} with $c=0$ plus $D$. Therefore,
\begin{align*}
\hat d(u,v)
&\le 2\min\left(
\sum_{i=1}^{\kappa}x_i,
\sum_{i=1}^{\kappa}y_i,
\min_{\substack{1\le i,j\le\kappa\\
x_i+y_j\ge 2D+3}}
\left\{
\sum_{r<i}x_r+\sum_{r<j}y_r+D
\right\}
\right)+D\\
&\le 2\kappa D+2\kappa+D \stackrel{\kappa=(k-1)/2}{=}
k\,D+k-1.
\end{align*}
This proves the theorem when $k$ is odd.

Now assume that $k$ is even. Then the $k-1$ levels cannot be split into two-level increments, and one level is left over. We pair the first level of one endpoint with the top level of the other endpoint. That is, we connect a vertex at distance about $h_1(u)$ from $u$ directly to $v$ through the top level of the hierarchy.

\begin{lemma}\label{lem:top-level-connection}
If $h_1(u)\le D+1$, then $\hat d(u,v)\le 2h_{k-2}(v)+3D+2-2h_1(u)$. The same holds with $u$ and $v$ exchanged.
\end{lemma}

\begin{proof}
Let $x$ be the vertex of $P(u,v)$ with $d(u,x)=\max\{h_1(u)-1,0\}$. Then $x\in B(u)$. Indeed, either $x=u$, or $d(u,x)+h_0(x)<h_1(u)$ and \Cref{lem:tz-bunch} gives $x=p_0(x)\in B(u)$. Moreover, $d(u,x)\ge h_1(u)-1$.

We connect $x$ to $v$ through the top level, as in \Cref{cl:pivot-connection}.
Suppose first that $d(v,x)+h_{k-2}(v)<h_{k-1}(x)$, and let $w=p_{k-2}(v)$. By \Cref{lem:tz-bunch}, $w\in B(x)$. Moreover,
$h_{k-1}(v)\ge h_{k-1}(x)-d(v,x)>h_{k-2}(v)$, so $w\in B(v)$, and $d(v,w)+d(w,x)\le 2h_{k-2}(v)+d(v,x)$.
Otherwise, $h_{k-1}(x)\le h_{k-2}(v)+d(v,x)$, and we let $w=p_{k-1}(x)$. Since $B_{k-1}(\cdot)=A_{k-1}$, we have $w\in B(v)\cap B(x)$, and
\[
d(v,w)+d(w,x)\le d(v,x)+2h_{k-1}(x)\le 2h_{k-2}(v)+3d(v,x).
\]

In both cases, $x,w\in B(x)$, so the algorithm ensures that $H(x,w)\le d(x,w)$. Since $x\in B(u)$ and $w\in B(v)$, the query gives
\begin{align*}
\hat d(u,v)
&\le d(u,x)+H(x,w)+d(w,v)\\
&\le d(u,x)+2h_{k-2}(v)+3d(x,v)\\
&=2h_{k-2}(v)+3D-2d(u,x)\\
&\le 2h_{k-2}(v)+3D+2-2h_1(u),
\end{align*}
as required.
\end{proof}

Let $\kappa=(k-2)/2$. We may assume that $h_1(u),h_1(v)\le D$. Indeed, if, say, $h_1(u)>D$, then $v\in B(u)$ by \Cref{lem:tz-bunch}, and the query returns $\hat d(u,v)=D$. We distinguish two cases.

\paragraph{Case 1: $h_1(u),h_1(v)\ge(D+1)/2$.}
Set $x_i=\nabla_{2i}(u)$ and $y_i=\nabla_{2i}(v)$ for $1\le i\le\kappa$. Then $\sum_{r<i}x_r=h_{2i-2}(u)$ and $\sum_{r<j}y_r=h_{2j-2}(v)$.
By \Cref{lem:top-level-connection} with $u$ and $v$ exchanged, and $h_1(v)\ge(D+1)/2$,
\[
\hat d(u,v)\le 2h_{k-2}(u)+2D+1=2\sum_{i=1}^{\kappa}x_i+2D+1,
\]
and symmetrically $\hat d(u,v)\le 2\sum_{i=1}^{\kappa}y_i+2D+1$.
By \Cref{lem:mixed-level-growth} with levels $2i$ and $2j$, if $x_i+y_j\ge2D+3$, then
\[
\hat d(u,v)\le 2\Bigl(\sum_{r<i}x_r+\sum_{r<j}y_r+D\Bigr)+D.
\]
Thus, $\hat d(u,v)$ is at most twice the optimization target of \Cref{lem:k-optimization} with $c=0$ plus $2D+1$, and
\[
\hat d(u,v)\le 2\kappa D+2\kappa+2D+1\stackrel{\kappa=(k-2)/2}{=}kD+k-1.
\]

\paragraph{Case 2: $h_1(u)<(D+1)/2$.}
The case $h_1(v)<(D+1)/2$ is symmetric. Here, we use the odd levels of $u$ and the even levels of $v$.
Set $x_i=\nabla_{2i+1}(u)$ and $y_i=\nabla_{2i}(v)$ for $1\le i\le\kappa$, and $c=(D+1)/2-h_1(u)>0$. Then
$\sum_{r<i}x_r=h_{2i-1}(u)-h_1(u)$ and $\sum_{r<j}y_r=h_{2j-2}(v)$.
The first bound above gives
\[
\hat d(u,v)\le 2h_{k-1}(u)+D=2\sum_{i=1}^{\kappa}x_i+2h_1(u)+D.
\]
By \Cref{lem:top-level-connection},
\[
\hat d(u,v)\le 2h_{k-2}(v)+3D+2-2h_1(u)=2\Bigl(\sum_{i=1}^{\kappa}y_i+2c\Bigr)+2h_1(u)+D.
\]
By \Cref{lem:mixed-level-growth} with levels $2i+1$ and $2j$, if $x_i+y_j\ge2D+3$, then
\[
\hat d(u,v)\le 2h_{2i-1}(u)+2h_{2j-2}(v)+3D=2\Bigl(\sum_{r<i}x_r+\sum_{r<j}y_r+D\Bigr)+2h_1(u)+D.
\]
Thus, by \Cref{lem:k-optimization},
\[
\hat d(u,v)\le 2(\kappa D+\kappa+c)+2h_1(u)+D=(2\kappa+2)D+2\kappa+1\stackrel{\kappa=(k-2)/2}{=}kD+k-1.
\]
This proves the theorem when $k$ is even.

Next, for completeness, we prove that the running time of the algorithm is $\Ot(mn^{1/k} + n^{1+2/k})$.
\begin{lemma}[Lemma~7 of~\cite{kadria2026improved}]\label{lem:k-alg-running-time}
    The algorithm takes $\Ot(mn^{1/k} + n^{1+2/k})$ time.
\end{lemma}
\begin{proof}
    Computing $B(u)$ for every $u \in V$ takes $\Ot(mn^{1/k})$ time.
    Constructing $H$ requires $O\left(\sum_{u \in V} |B(u)|^2\right) = \Ot(n^{1 + 2/k})$ time.
    Estimating the distances takes $O\left(\sum_{\pair{u,v} \in I} |B(u)| \cdot |B(v)|\right) = \Ot(n^{1 + 2/k})$ time. Therefore, the running time is $\Ot(mn^{1/k} + n^{1+2/k})$, as required.
\end{proof}

\paragraph{Lower bound for the $k$-stretch algorithm.}
The factor $k$ is tight for this algorithm: consider a pair $u,v$ at distance $D$ for which the relevant pivot distances increase by $D/2$ at each level. The hierarchy can be arranged so that the strict bunch inequalities are never satisfied along $P(u,v)$ before the last level, and hence no shorter route is available through a connecting vertex $x$ at any level. 
At level $k-1$, the first available route can have length $D+2h_{k-1}(x)=D+2(k-1)D/2=kD$.

\section{$\frac 65 k$-stretch in $\widetilde{O}(mn^{1/k})$ time}
\label{sec:65k}

In this section, we provide a \((\frac 65 k+O(1))\)-multiplicative approximation algorithm for \mPSP that runs in \(\Ot(mn^{1/k})\) time, for both weighted and unweighted graphs. We remove the time-consuming parts of \Cref{alg:k-alg} and instead use (only) a preprocessing technique from~\cite{kadria2026improved}, originally introduced to obtain a \(\sim 1.622k\)-approximation algorithm for \nPSP on weighted graphs.

\Reminder{thm:intro-weighted}

Throughout this section, let $I$ be the set of $m$ queried vertex pairs.

We first explain which parts of \Cref{alg:k-alg} cause its high running time and why its stretch analysis does not extend to weighted graphs, and describe how~\cite{kadria2026improved} overcame the latter issue and obtained a stretch better than \(2k-1\).

\subsection{The problems (and some of the solutions) of \Cref{alg:k-alg}}

When \(m=o(n^{1+1/k})\), the running time of \Cref{alg:k-alg} is dominated by the \(\Ot(n^{1+2/k})\) term. This term arises in two places. First, during preprocessing, we store in \(H\) the distances between every pair of vertices in \(B(x)\), for every \(x\in V\). These distances are used when the estimate is obtained through \(p_i(u),p_j(v)\in B(x)\):
\[
\hat d(u,v)\le d(u,p_i(u))+H(p_i(u),p_j(v))+d(p_j(v),v).
\]

Second, during the query, we examine every pair in \(B(u)\times B(v)\). This is needed when we find \(p_i(x)\in B(u)\) and \(p_j(x)\in B(v)\), giving
\[
\hat d(u,v)\le d(u,p_i(x))+H(p_i(x),p_j(x))+d(p_j(x),v).
\]

This algorithm also fails to achieve the same stretch guarantee in weighted graphs. In \Cref{lem:mixed-level-growth}, the additive \(3\) accounts for rounding the ideal split point on the path $P(u,v)$ to a nearby vertex. In a weighted graph, however, a single edge of \(P(u,v)\) may contain almost the entire distance \(d(u,v)\), so this rounding can incur an additive \(\Theta(d(u,v))\) loss at each relevant step. Consequently, the stretch deteriorates back to \(2k-1\).

Kadria et al.~\cite{kadria2026improved} improved the stretch by augmenting the preprocessing as follows: for every oriented edge \((x,y)\) and every \(i\in[k]\), they also stored \(H(p_i(x),z)\) for every \(z\in B(y)\). Over all edges, this requires only \(\Ot(mn^{1/k})\) time.

If \((x,y)\) is a heavy edge on \(P(u,v)\), this preprocessing stores \(H(p_i(u),p_j(y))\) whenever \(p_i(u)\in B(x)\), \(H(p_i(x),p_j(v))\) whenever \(p_j(v)\in B(y)\), and \(H(p_i(x),p_j(y))\) for every \(i,j\in[k]\). In all three cases, the connection can cross the heavy edge directly through the stored entry in \(H\), instead of forcing the analysis to round the split point to either \(x\) or \(y\) and pay for the edge weight. However, the preprocessing does not provide \(H(p_i(u),p_j(v))\). This remaining case prevents the weighted analysis of~\cite{kadria2026improved} from matching the \(\frac{4}{3}k\) stretch obtained for unweighted graphs, resulting instead in a stretch of approximately \(1.622k\).

\subsection{Our algorithm}

We show that a stretch of \(\frac 65 k+O(1)\) can be guaranteed, and our analysis uses only estimates of the forms
\[
u\to p_i(u)
\xrightarrow[\text{via }x\to y]{H}
p_j(y)\to v
\qquad\text{and}\qquad
u\to p_i(x)
\xrightarrow[\text{via }x\to y]{H}
p_j(v)\to v,
\]
where \((x,y)\) is an edge of $P(u,v)$ and \(i,j\in[k]\). The query may find other walks as well, but every estimate is the length of some $u$--$v$ walk, so $\hat d(u,v)\ge d(u,v)$. Restricting the analysis to these two forms helps us in three ways:
\begin{itemize}
    \item It eliminates the preprocessing over all pairs in \(B(x)\times B(x)\), removing the \(\Ot(n^{1+2/k})\) term.
    \item It replaces the search over \(B(u)\times B(v)\) by considering \(\{p_i(u)\mid i\in[k]\}\times B(v)\) and \(B(u)\times\{p_j(v)\mid j\in[k]\}\), which takes \(\Ot(n^{1/k})\) time per pair. Consequently, the algorithm solves \mPSP in \(\Ot(mn^{1/k})\) time.
    \item It allows us to incorporate the weighted-graph preprocessing technique of~\cite{kadria2026improved} without any additional loss in the stretch.
\end{itemize}

The algorithm works as follows. It first initializes two empty hash tables \(H\) and \(\hat d\), and computes \(B(u)\) for every \(u\in V\) using \Cref{lem:tz-construction}.\footnote{As in \Cref{alg:k-alg}, \(H\) returns \(\infty\) on unset pairs, and \(H(w,w)=0\) for every \(w\).} Then, for every oriented edge \((x,y)\), every \(i\in[k]\), and every \(z\in B(y)\), it sets \(H(p_i(x),z)\) and \(H(z,p_i(x))\) to \(\min\{H(p_i(x),z),h_i(x)+\wt(x,y)+d(y,z)\}\). Next, for every pair \(\pair{u,v}\in I\), the algorithm considers two types of estimates. For every \(i\in[k]\) and \(w\in B(v)\), it sets \(\hat d(u,v)\) to \(\min\{\hat d(u,v),d(u,p_i(u))+H(p_i(u),w)+d(w,v)\}\). Symmetrically, for every \(z\in B(u)\) and \(j\in[k]\), it sets \(\hat d(u,v)\) to \(\min\{\hat d(u,v),d(u,z)+H(z,p_j(v))+d(p_j(v),v)\}\). Finally, the algorithm returns \(\hat d(u,v)\) for every \(\pair{u,v}\in I\). Pseudocode is presented in \Cref{alg:65k-alg}. See \Cref{fig:lb-estimate} for an illustration of a possible estimate.

\begin{algorithm2e}
    \caption{\(\frac 65 k\)-approximation for \mPSP}
    \label{alg:65k-alg}

    \(H\gets\HashTable();\ \hat d\gets\HashTable()\)\;
    Compute \(B(u)\) for every \(u\in V\) using \Cref{lem:tz-construction}\;

    \ForEach{oriented edge \((x,y)\)}{
        \For{\(i\in[k]\) and \(z\in B(y)\)}{
            \(H(p_i(x),z), H(z,p_i(x))\gets
            \min\{H(p_i(x),z),h_i(x)+\wt(x,y)+d(y,z)\}\)\;
        }
    }

    \ForEach{\(\pair{u,v}\in I\)}{
        \For{\(i\in[k]\) and \(w\in B(v)\)}{
            \(\hat d(u,v)\gets
            \min\{\hat d(u,v),
            d(u,p_i(u))+H(p_i(u),w)+d(w,v)\}\)\;
        }

        \For{\(z\in B(u)\) and \(j\in[k]\)}{
            \(\hat d(u,v)\gets
            \min\{\hat d(u,v),
            d(u,z)+H(z,p_j(v))+d(p_j(v),v)\}\)\;
        }
    }

    \Return \(\{\hat d(u,v)\mid \pair{u,v}\in I\}\)\;
\end{algorithm2e}

To obtain the improved approximation guarantee, we develop two new analytical tools for the Thorup--Zwick
hierarchy, formalized in the \emph{Two-Level Growth Lemma} and the \emph{Long--Short Growth Lemma}, that translate large increases of the pivot distances \(h_i(u)\) and \(h_i(v)\) into short query paths. We describe these lemmas next. Together, they reduce the approximation analysis to a numerical optimization problem whose solution yields the improved stretch.

For the rest of this section, fix a pair \(u,v\in V\), and let \(P=P(u,v)\) and \(D=d(u,v)\). Recall the one-level growth \(\Delta_i(x)=h_i(x)-h_{i-1}(x)\) and the two-level growth \(\nabla_i(x)=h_i(x)-h_{i-2}(x)=\Delta_i(x)+\Delta_{i-1}(x)\) from \Cref{sec:prelim}.

\subsection{The Two-Level Growth Lemma}
\begin{lemma}[Two-Level Growth Lemma]
\label{lem:two-level-growth}
If for some $2 \le i < k$,
\[
h_i(u) - h_{i-2}(v) > D,
\]
then
\[
\hat{d}(u,v)
\le 2h_{i-2}(v)+h_{i-1}(u)+h_i(u)+D.
\]
This lemma holds symmetrically when switching between $u$ and $v$.
\end{lemma}

Intuitively, we use the intermediate level $i-1$ to connect level $i$ at $u$ to level $i-2$ at $v$. Notice that for every $z\in P(u,v)$,
\[
h_i(u)-h_{i-2}(v)=\bigl(h_i(u)-h_{i-1}(z)\bigr)
+
\bigl(h_{i-1}(z)-h_{i-2}(v)\bigr)>D.
\]
Since $d(u,z)+d(z,v)=D$, either $\bigl(h_i(u)-h_{i-1}(z)\bigr)>d(u,z)$ or $\bigl(h_{i-1}(z)-h_{i-2}(v)\bigr)>d(z,v)$ holds. Moving along the path, we find an edge $(x,y)$ such that 
\[
h_i(u)-h_{i-1}(x)>d(u,x)
\qquad\text{and}\qquad
h_{i-1}(y)-h_{i-2}(v) > d(y,v),
\]
so
\[
p_{i-1}(x)\in B(u)
\qquad\text{and}\qquad
p_{i-2}(v)\in B(y).
\]
Processing this edge stores a connection between $p_{i-1}(x)$ and $p_{i-2}(v)$, which the query uses to connect $u$ to $v$.

\begin{proof}
We first handle the cases in which a pivot of $u$ or $v$ suffices. If $h_{i-1}(v)=h_{i-2}(v)$, then
\[
h_i(u)>h_{i-2}(v)+D=h_{i-1}(v)+D.
\]
By \Cref{lem:tz-bunch}, $p_{i-1}(v)\in B(u)$, and the direct query gives
\[
\hat d(u,v)
\le D+2h_{i-1}(v)
=D+2h_{i-2}(v),
\]
which is stronger than the required bound. If $h_{i-1}(u)>h_{i-2}(v)+D$, then
\Cref{lem:tz-bunch} gives $p_{i-2}(v)\in B(u)$, and again
\[
\hat d(u,v)\le D+2h_{i-2}(v).
\]

We may therefore assume that
\[
h_{i-1}(v)>h_{i-2}(v)
\qquad\text{and}\qquad
h_{i-1}(u)\le h_{i-2}(v)+D.
\]
As in the proof of \Cref{lem:mixed-level-growth}, we now
find a connection along $P(u,v)$. Here, the connection uses an edge
rather than a single vertex.

\begin{cclaim}\label{cl:two-level-connecting-edge}
There exists an edge $(x,y)$ of $P(u,v)$, oriented from $u$ to $v$,
such that
\[
h_i(u)-h_{i-1}(x)>d(u,x)
\qquad\text{and}\qquad
h_{i-1}(y)-h_{i-2}(v)>d(y,v).
\]
\end{cclaim}

\begin{proof}
For every vertex $q\in P(u,v)$, define
\[
F(q)=h_i(u)-h_{i-1}(q)-d(u,q),
\qquad
G(q)=h_{i-1}(q)-h_{i-2}(v)-d(q,v).
\]
Our goal is to find an edge $(x,y)$ with $F(x)>0$ and $G(y)>0$.
For every $q\in P(u,v)$,
\[
F(q)+G(q)=h_i(u)-h_{i-2}(v)-D>0.
\]
Moreover, our assumptions give $G(u)\le0$ and $G(v)>0$.

Let $y$ be the first vertex of $P(u,v)$, starting from $u$,
such that $G(y)>0$, and let $x$ be its predecessor.
Then $G(x)\le0$, so
\[
F(x)>-G(x)\ge0.
\]
Thus, $(x,y)$ satisfies both required inequalities.
\end{proof}

By \Cref{cl:two-level-connecting-edge} and \Cref{lem:tz-bunch},
\[
p_{i-1}(x)\in B(u)
\qquad\text{and}\qquad
p_{i-2}(v)\in B(y).
\]
Therefore, when processing the edge $(x,y)$, the algorithm ensures that
\[
H\bigl(p_{i-1}(x),p_{i-2}(v)\bigr)
\le h_{i-1}(x)+d(x,y)+d\bigl(y,p_{i-2}(v)\bigr),
\]
where we used that the weight of $(x,y)$ equals $d(x,y)$,
since it lies on a shortest path.

The query considers this pair of pivots and gives
\begin{align*}
\hat d(u,v)
&\le d\bigl(u,p_{i-1}(x)\bigr)
 +H\bigl(p_{i-1}(x),p_{i-2}(v)\bigr)
 +h_{i-2}(v)\\
&\le d(u,x)+2h_{i-1}(x)+d(x,y)+d(y,v)+2h_{i-2}(v)\\
&=D+2h_{i-1}(x)+2h_{i-2}(v).
\end{align*}

Finally, \Cref{cl:two-level-connecting-edge} and the triangle inequality
give, respectively,
\[
h_{i-1}(x)<h_i(u)-d(u,x),
\qquad
h_{i-1}(x)\le h_{i-1}(u)+d(u,x).
\]
Adding these inequalities yields
\[
2h_{i-1}(x)\le h_i(u)+h_{i-1}(u).
\]
Substituting into the bound above, we obtain
\[
\hat d(u,v)\le
2h_{i-2}(v)+h_{i-1}(u)+h_i(u)+D,
\]
as required.
\end{proof}

\subsection{The Long--Short Growth Lemma}

\begin{lemma}[Long--Short Growth Lemma]
\label{lem:long-short} If for some $2 \le i <k$ and $1 \le j < k$,
\[
\nabla_i(u)+\Delta_j(v)>2D,
\] 
then 
\[
\hat{d}(u,v) \le 2h_{i-1}(u)+2h_{j-1}(v) + 3D.
\] 
This lemma holds symmetrically when switching between $u$ and $v$. 
\end{lemma}

Intuitively, the argument works as follows. We choose \(x \in P(u,v)\) close to \(v\), so that the subpath from \(u\) to \(x\) contains almost all of \(P(u,v)\).
We first handle this long subpath. If
\[
h_i(u)-h_{i-2}(u)>2d(u,x),
\]
then one of two possibilities must hold: either a pivot of \(u\) belongs to \(B(x)\), or a pivot of \(x\) belongs to \(B(u)\). At this stage, we deliberately keep both possibilities open. We only care that the long part of the path is covered, not whether it is covered through a pivot of \(u\) or through a pivot of \(x\).

Once this choice is made, the short subpath from \(x\) to \(v\) must complete the corresponding connection. If the long part uses a pivot of \(u\), we need a pivot of \(x\) that belongs to \(B(v)\). If the long part uses a pivot of \(x\), we instead need a pivot of \(v\) that belongs to \(B(x)\).
The one-level gap
\[
h_j(v)-h_{j-1}(v)>2d(v,x)
\]
guarantees both possibilities. Thus, the short part can adapt to whichever choice was made for the long part. This is similar to a Thorup--Zwick routing argument between \(x\) and \(v\): the large gap guarantees that the appropriate previous-level pivot of one endpoint belongs to the bunch of the other endpoint.

Since
\(
d(u,x)+d(v,x)=D,
\)
it is enough that
\[
\bigl(h_i(u)-h_{i-2}(u)\bigr)
+
\bigl(h_j(v)-h_{j-1}(v)\bigr) > 2D.
\]

Moreover, by adding \(H(p_i(x),z)\) for every edge \((x,y)\in E\), every level \(i\), and every \(z\in B(y)\), we can split the path across the edge \((x,y)\), and check whether
\[
h_i(u)-h_{i-2}(u)>2d(u,x) \qquad\text{and}\qquad h_j(v)-h_{j-1}(v)>2d(y,v).
\]
This is particularly useful in weighted graphs, where the ideal split point may lie inside a heavy edge. Without this preprocessing, we would have to move the split point to either \(x\) or \(y\), forcing one side of the analysis to include the full weight of the edge. Instead, the first part only needs to reach \(x\), while the second part starts from \(y\), and the stored connection crosses the edge directly. Thus, we do not need to pay for the weight of \((x,y)\) in either of the two gaps.

\vspace{5pt}
We now prove the Long--Short Growth Lemma. We assume that \(\nabla_i(u),\Delta_j(v)\le 2D\).\footnote{If \(\nabla_i(u)>2D\), then \Cref{lem:tz-bunch} gives \(p_{i-2}(u)\in B(v)\) or \(p_{i-1}(v)\in B(u)\). If \(\Delta_j(v)>2D\), then \Cref{lem:one-level-gap} gives \(p_{j-1}(v)\in B(u)\). In each case, the direct query already gives the required bound.} We split the rest of the proof into two claims.
\begin{cclaim}[Long--Short Growth Lemma Part 1] \label{cl:long-short-split-edge}
If for some $2 \le i <k$ and $1 \le j < k$,
\[ \nabla_i(u)+\Delta_j(v)>2D \quad\text{and}\quad \nabla_i(u),\Delta_j(v)\le 2D, \]
then there exists an edge \((x,y)\) of \(P\), oriented so that \(u,x,y,v\) appear in this order, such that 
\[ \nabla_i(u)>2d(u,x) \quad \text{and} \quad \Delta_j(v)>2d(y,v).
\]
\end{cclaim}

\begin{proof}
See \Cref{fig:long-short} for an illustration. We view each edge of \(P\) as a continuous interval whose length equals its weight. Thus, every value in \([0,D]\) is the distance \(d(u,z)\) of some point \(z\) on \(P\).

By the assumptions,
\[
0\le D-\frac{\Delta_j(v)}{2}
<
\frac{\nabla_i(u)}{2}\le D.
\]
We can choose a point \(z\) on the continuous path \(P\) such that
\(D-\Delta_j(v)/2<d(u,z)<\nabla_i(u)/2\).
Note that \(z\) is not necessarily a vertex of \(G\): it may be an imaginary point in the interior of an edge, and it is used only in the analysis, not by the algorithm.

Let \((x,y)\) be the edge containing \(z\), oriented so that
\(u,x,z,y,v\) appear in this order. Since \(x\) appears before \(z\),
\[
2d(u,x)\le 2d(u,z)<\nabla_i(u).
\]
Similarly, since \(y\) appears after \(z\), we have
\(d(y,v)\le d(z,v)=D-d(u,z)<\Delta_j(v)/2\), and therefore
\(
2d(y,v)<\Delta_j(v).
\)
Thus, \((x,y)\) satisfies both required inequalities.
\end{proof}

\begin{cclaim}[Long--Short Growth Lemma Part 2] \label{cl:long-short-mixed-connection}
Let \((x,y)\) be an edge of \(P(u,v)\), oriented so that
\(u,x,y,v\) appear in this order. Suppose that there are $2 \le i <k$ and $1 \le j < k$ such that
\[
\nabla_i(u)>2d(u,x) \quad \text{ and } \quad \Delta_j(v)>2d(y,v).
\]
Then
\[
\hat d(u,v)
\le
2h_{i-1}(u)+2h_{j-1}(v) + 3D.
\]
\end{cclaim}

\begin{proof}
We first use the two-level gap at \(u\). We have
\[
\begin{aligned}
h_i(u)-h_{i-2}(u)
={}
\bigl(h_{i-1}(x)-h_{i-2}(u)\bigr) +
\bigl(h_i(u)-h_{i-1}(x)\bigr).
\end{aligned}
\]
Since this quantity is greater than \(2d(u,x)\), at least one of the following inequalities holds:
\[
h_{i-1}(x)-h_{i-2}(u)>d(u,x) \quad \text{or} \quad h_i(u)-h_{i-1}(x)>d(u,x).
\]

By \Cref{lem:tz-bunch}, in the first case
\(
p_{i-2}(u)\in B(x),
\)
and in the second case
\(
p_{i-1}(x)\in B(u).
\)

We next use the one-level gap at \(v\). Since
\(
h_j(v)-h_{j-1}(v)>2d(y,v),
\)
\Cref{lem:one-level-gap} gives both
\(
p_{j-1}(y)\in B(v)
\)
and
\(
p_{j-1}(v)\in B(y).
\)

Thus, if \(p_{i-1}(x)\in B(u)\), the preprocessing for the edge \((x,y)\) lets the algorithm consider the walk
\[
u \to p_{i-1}(x)\to x\to y\to p_{j-1}(v) \to v.
\]
In this case,
\[
d\bigl(u,p_{i-1}(x)\bigr)+d\bigl(p_{i-1}(x),x\bigr)
\le 2h_{i-1}(x)+d(u,x) \le 2h_{i-1}(u) + 3d(u,x),
\]
where in the last inequality we used the fact that $h_{i-1}(x) \le h_{i-1}(u) + d(u,x)$. Moreover,
\[
\begin{aligned}
d\bigl(y,p_{j-1}(v)\bigr)+d\bigl(p_{j-1}(v),v\bigr) \le d(y,v) + 2h_{j-1}(v).
\end{aligned}
\]
Therefore,
\begin{align*}
\hat d(u,v)
&\le
d\bigl(u,p_{i-1}(x)\bigr)
+H\bigl(p_{i-1}(x),p_{j-1}(v)\bigr)
+d\bigl(p_{j-1}(v),v\bigr)\\
&\le
d\bigl(u,p_{i-1}(x)\bigr)
+d\bigl(p_{i-1}(x),x\bigr)
+\wt(x,y)
+d\bigl(y,p_{j-1}(v)\bigr)
+d\bigl(p_{j-1}(v),v\bigr)\\
&\le
2h_{i-1}(u) + 3d(u,x)
+\wt(x,y) +d(y,v) + 2h_{j-1}(v)\\
&\le 2h_{i-1}(u)+2h_{j-1}(v)+3D.
\end{align*}

The case where the query uses \(p_{i-2}(u)\in B(x)\) and \(p_{j-1}(y)\in B(v)\) follows similarly and yields a slightly stronger bound.

\end{proof}

\subsection{The optimization}
\label{sec:65k-optimization}

Next, we use both the Two-Level Growth Lemma and the Long--Short Growth Lemma to obtain a good approximation ratio for the algorithm. We have the following three bounds.
\begin{enumerate}
    \item $\hat{d}(u,v) \leq 2\min(h_{k-1}(u),h_{k-1}(v)) + D$.
    \item By \Cref{lem:two-level-growth}, if $h_i(u)>h_{i-2}(v)+D$, then $\hat{d}(u,v)\le 2h_{i-2}(v)+h_{i-1}(u)+h_i(u)+D$. This holds symmetrically when switching between $u$ and $v$.
    \item By \Cref{lem:long-short}, if $\nabla_i(u)+\Delta_j(v)>2D$, then $\hat{d}(u,v)\le 2h_{i-1}(u)+2h_{j-1}(v) + 3D$. This holds symmetrically when switching between $u$ and $v$.
\end{enumerate}
At this point, all graph-theoretic aspects of the analysis are captured by the bounds above. It remains to show that at least one of these bounds is sufficiently small.

We apply the following lemma with $x_i=\Delta_i(u)/D$ and $y_i=\Delta_i(v)/D$, so that $X_i=h_i(u)/D$ and $Y_i=h_i(v)/D$.\footnote{
We may assume $x_1,y_1\le2$, since otherwise a direct level-zero query already returns the exact distance $D$.}
\begin{lemma}
\label{lem:six-fifths-optimization}\Copy{lem:six-fifths-optimization}{
Let \(x_1,\ldots,x_{k-1},y_1,\ldots,y_{k-1}\ge0\) with \(x_1,y_1\le 2\), and for every \(0\le i\le k-1\) let
\(
X_i=\sum_{r=1}^{i}x_r
\)
and
\(
Y_i=\sum_{r=1}^{i}y_r.
\)
Then
\begin{align*}
\min\Bigg\{
&X_{k-1},\ Y_{k-1},
\\
&\min_{\substack{2\le i\le k-1\\
X_i>Y_{i-2}+1}}
\left(Y_{i-2}+\frac{X_{i-1}+X_i}{2}\right),
\min_{\substack{2\le i\le k-1\\
Y_i>X_{i-2}+1}}
\left(X_{i-2}+\frac{Y_{i-1}+Y_i}{2}\right),
\\
&\min_{\substack{2\le i\le k-1,\ 1\le j\le k-1\\
x_i+x_{i-1}+y_j>2}}
\left(X_{i-1}+Y_{j-1}\right),
\min_{\substack{2\le i\le k-1,\ 1\le j\le k-1\\
y_i+y_{i-1}+x_j>2}}
\left(Y_{i-1}+X_{j-1}\right)
\Bigg\}
\le
\frac{3}{5}k+2.
\end{align*}
}
\end{lemma}

We defer the proof of \Cref{lem:six-fifths-optimization} to \Cref{app:six-fifths-optimization}. Let $\sol$ denote the value of the minimum in the lemma. By the definitions of $X_i,Y_i,x_i$, and $y_i$, every feasible expression in this minimum gives, through the corresponding bound above, an estimate of at most $2D$ times that expression plus $3D$. Therefore,
\[
\hat{d}(u,v)
\leq
2D\cdot\sol+3D
\leq
\left(\frac 65 k+7\right)D.
\]

\begin{figure}[t]
    \centering
\begin{tikzpicture}[x=1cm,y=0.8cm,line width=0.75pt,
    every node/.style={font=\small,inner sep=1.5pt},
    stem/.style={densely dashed,line width=0.6pt},
    lvl/.style={densely dotted,line width=0.5pt,gray}]

  \draw (0,0) -- (5,0);
  \foreach \t in {1,...,4} \fill (\t,0) circle (1.3pt);
  \fill (0,0) circle (2.2pt);
  \fill (5,0) circle (2.2pt);
  \node[below=2pt] at (0,0) {$u$};
  \node[below=2pt] at (5,0) {$v$};
  \node[below=2pt,font=\footnotesize] at (2.5,0) {$D$};

  \draw[stem] (0,0) -- (0,5.25);
  \draw[stem] (5,0) -- (5,5.25);
  \foreach \i in {1,...,6} {
    \fill (0,{\i/2}) circle (1.6pt);
    \fill (5,{\i/2}) circle (1.6pt);
  }
  \foreach \i in {7,...,9} {
    \fill (0,{3+0.75*(\i-6)}) circle (1.6pt);
    \fill (5,{3+0.75*(\i-6)}) circle (1.6pt);
  }

  \draw[lvl] (0,3) -- (5,3);
  \draw[lvl] (0,5.25) -- (5,5.25);
  \node[left=4pt,font=\footnotesize] at (0,0.5) {$p_1(u)$};
  \node[left=4pt,font=\footnotesize] at (0,3) {$p_q(u)$};
  \node[left=4pt,font=\footnotesize] at (0,5.25) {$p_{k-1}(u)$};
  \node[right=4pt,font=\footnotesize] at (5,0.5) {$p_1(v)$};
  \node[right=4pt,font=\footnotesize] at (5,3) {$p_q(v)$};
  \node[right=4pt,font=\footnotesize] at (5,5.25) {$p_{k-1}(v)$};

  \node[font=\footnotesize,align=center] at (2.5,1.6)
      {$h_i$ grows by $D/2$ per level};
  \node[font=\footnotesize,align=center] at (2.5,4.15)
      {$h_i$ grows by $3D/4$ per level};
\end{tikzpicture}

    \caption{The pivot distances $h_i(u)=h_i(v)$ that are tight for
    \Cref{lem:six-fifths-optimization}, with $q\approx\frac35k$. Both
    $2h_{k-1}(u)$ and $2h_q(u)+2h_q(v)$ are $\approx\frac 65 kD$.}
    \label{fig:growth-profile}
\end{figure}
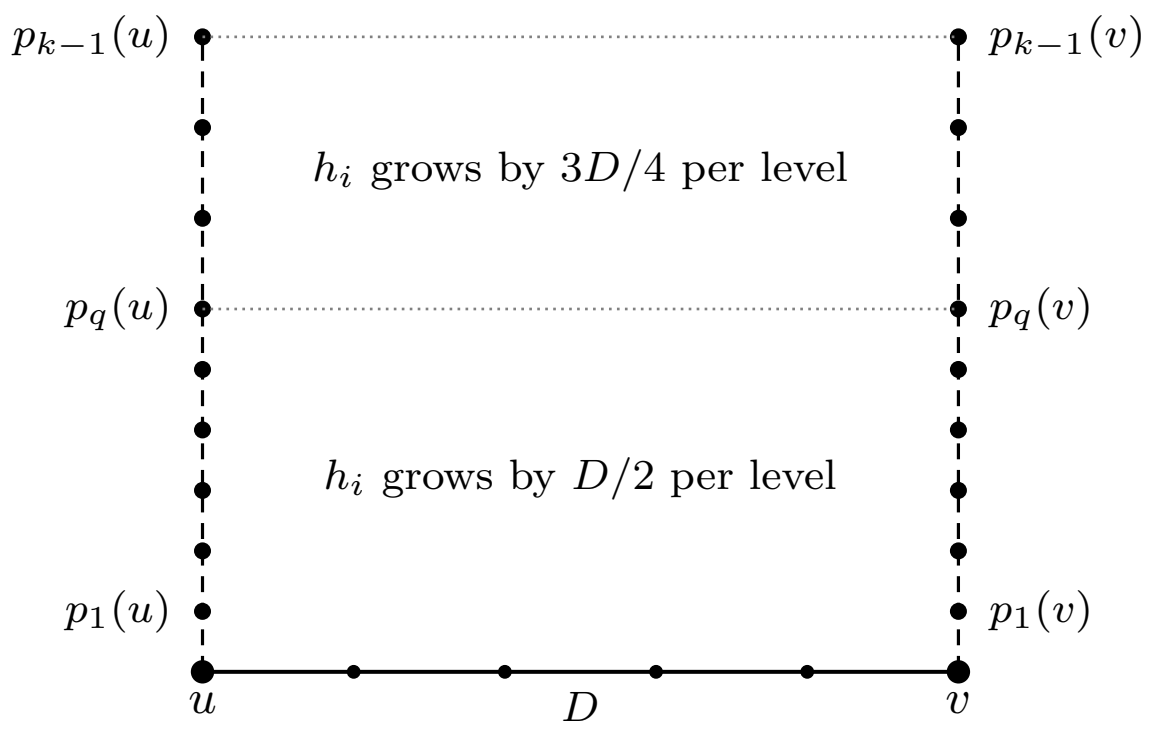

Although the proof is rather long and technical, we illustrate the main idea under the simplifying assumption that $h_i(u)=h_i(v)$ for every $i$ (see \Cref{fig:growth-profile}). In this case, $x_i=y_i$ and $X_i=Y_i$ for every $i$, and the optimization problem reduces to showing that
\begin{align*}
\min\Bigg\{
X_{k-1}, \min_{\substack{2\le i\le k-1\\
x_i+x_{i-1}>1}}
\left(X_{i-2}+\frac{X_{i-1}+X_i}{2}\right),
\min_{\substack{2\le i\le k-1,\ 1\le j\le k-1\\
x_i+x_{i-1}+x_j>2}}
\left(X_{i-1}+X_{j-1}\right)
\Bigg\}
\le
\frac{3}{5}k+2.
\end{align*}

Let $\sol$ denote the value of this optimization problem, and let $q$ be the largest index such that $2X_q<\sol$. By the maximality of $q$, we have $2X_{q+1}\geq\sol$, so $2X_q \ge \sol -2$. Moreover, for every $2\le i\le q$,
\[
X_{i-2}+\frac{X_{i-1}+X_i}{2}
\le 2X_i
\le 2X_q
<\sol.
\]
Therefore, the two-level term implies $x_i+x_{i-1}\le1$. Summing over disjoint pairs of consecutive increments gives $2X_q\le q+O(1)$.

To illustrate the main idea, we ignore additive constants and think of the transition as occurring around $q\approx\sol$ and $X_q\approx\sol/2$. Thus, the average value of the increments $x_1,\ldots,x_q$ is approximately $1/2$. Suppose, for simplicity, that all these increments are exactly $1/2$.

Now consider a later level $i > q$. As long as $X_{i-1}+X_{j-1}<\sol$, the long--short term in the optimization problem implies
\(
x_i+x_{i-1}+x_j\leq 2.
\)
Taking any $j\leq q$, for which $x_j=1/2$, gives
\(
x_i+x_{i-1}\leq\frac{3}{2}.
\)
Hence, the average increase after level $q$ is at most $3/4$. The first $q$ levels contribute approximately $q/2$ to $X_{k-1}$, while the remaining $k-q$ levels contribute at most approximately $3(k-q)/4$. Since $\sol \leq X_{k-1}$, we obtain
\[
\sol \le X_{k-1}
\lesssim
\frac{q}{2}+\frac{3}{4}(k-q).
\]
Moreover, we have $q\approx\sol$. Substituting this relation gives
\(
\sol
\lesssim
\frac{\sol}{2}+\frac{3}{4}(k-\sol),
\)
which rearranges to $\sol\lesssim 3k/5$.

\vspace{0.5\baselineskip}

Lastly, we prove that the running time of the algorithm is $\Ot(mn^{1/k})$ and complete the proof of \Cref{thm:intro-weighted}.
\begin{lemma}\label{lem:weighted-running-time}
    \Cref{alg:65k-alg} runs in $\Ot(mn^{1/k})$ time.
\end{lemma}
\begin{proof}
    Computing $B(u)$ for every $u \in V$ takes $\Ot(mn^{1/k})$ time.
    Constructing $H$ requires, for every edge $(x,y)\in E$ and every $i\in[k]$, computing the values $h_i(x) + \wt(x,y) + d(y,z)$ for every $z \in B(y)$. For every edge, this takes $\Ot(n^{1/k})$ time, so for all edges this takes $\Ot(mn^{1/k})$ time.
    Estimating the distances takes 
    \[
    O\left(k \sum_{\pair{u,v} \in I} \bigl(|B(u)|+|B(v)|\bigr)\right) = \Ot(m n^{1/k}).
    \]
     Therefore, the running time is $\Ot(mn^{1/k})$, as required.
\end{proof}

\paragraph{Lower bound for the $6/5$ constant.}
Let $D=d(u,v)$ and let $q\approx\frac35k$ be an integer. In our lower-bound
instance, $u$ and $v$ are symmetric, and their pivot distances follow a
two-phase profile: $h_i(u)=h_i(v)$ grows by $\frac D2$ per level up to
level $q$, and by $\frac34D$ per level afterwards (see \Cref{fig:growth-profile}). This is the fastest
profile that the Two-Level Growth Lemma and the Long--Short Growth Lemma
permit, the first phase being bounded by the
former and the second by the latter.

In this hierarchy, every walk considered by \Cref{alg:65k-alg} incurs
at least one of the following two costs.
\begin{itemize}
\item Walks through a pivot of the last level $A_{k-1}$, whose distance
$h_{k-1}(u)\approx\frac q2D+\frac34(k-1-q)D$ to $P(u,v)$ is
therefore traversed twice:
\[
\hat{d}(u,v)\ =\ D+2h_{k-1}(u)
\ \approx\ 2\left(\tfrac q2D+\tfrac34(k-1-q)D\right)\ \approx\ \frac 65 kD .
\]
\item Walks through two separate pivots of level about $q$.
The bunch and pivot conditions force both pivots to be at levels at least $q-1$.
Each is about $\frac q2 D$ from the path, and each detour is traversed twice:
\[
\hat{d}(u,v)\ =\ D+2h_q(u)+2h_q(v)\ \ge\ 2\frac q2 D+2\frac q2 D\ \approx\ \frac 65 kD.
\]
\end{itemize}
Thus, both costs give stretch at least $\frac65 k-O(1)$.

\section{Achieving improved stretch for short paths}
\label{sec:short-paths}
In this section, we show that the algorithm of \Cref{sec:65k}
achieves stretch $k$ on pairs whose shortest path is short, a case left open by the lower-bound discussions in \Cref{sec:k-stretch,sec:65k}.
We consider weighted graphs and pairs whose shortest path has at most two edges.

\Reminder{thm:nPSP-2-hop}
\begin{proof}
    Let $I$ be the set of $m$ queried vertex pairs, where $P(u,v)$ has at most two edges for every $\pair{u,v}\in I$.
    We run \Cref{alg:65k-alg} with parameter $k$ and choose $p_0(z)=z$ for
    every $z\in V$. We evaluate each input pair in both endpoint orders and
    take the smaller estimate. This changes the running time by only a
    constant factor, so the running-time bound follows immediately from
    \Cref{lem:weighted-running-time}. It remains to prove the stretch guarantee.

    Fix $\pair{u,v}\in I$ and assume that $P(u,v)$ has exactly two edges.
    Write $P(u,v)=(u,x,v)$. By symmetry between $u$ and $v$, assume that
    $\ell(u,x)\ge\ell(x,v)$. In particular,
    $\ell(x,v)\le d(u,v)/2$.

    Let $i\in\{1,\ldots,k\}$ be the smallest index such that
    $h_i(v)>i\ell(x,v)$ or $h_i(x)>i\ell(x,v)$. Such an index exists since
    $h_k(v)=h_k(x)=\infty$. We show that in either case, the additive error is at most
    $2(i-1)\ell(x,v)$.

    By the minimality of $i$, we have
    \begin{equation}\label{eq:small-hop-h-i-1}
        h_{i-1}(v),h_{i-1}(x)\le(i-1)\ell(x,v).
    \end{equation}

    If $h_i(v)>i\ell(x,v)$, then
    \[
        d(v,p_{i-1}(x))
        \stactri\le \ell(x,v)+h_{i-1}(x)
        \stackrel{\eqref{eq:small-hop-h-i-1}}\le \ell(x,v)+(i-1)\ell(x,v)=i\ell(x,v)
        < h_i(v),
    \]
    and therefore $p_{i-1}(x)\in B(v)$.
    Moreover, if $i=k$, then $h_{i-1}(x)<h_i(x)=\infty$, and hence
    $p_{i-1}(x)\in B_{i-1}(x)\subseteq B(x)$.
    Otherwise, the triangle inequality gives
    $h_i(v)\le h_i(x)+\ell(x,v)$, and hence
    \[
        h_i(x)>(i-1)\ell(x,v)\ge h_{i-1}(x),
    \]
    so again $p_{i-1}(x)\in B(x)$. Therefore, the query for $(u,v)$ gives
    \begin{align*}
        \hat d(u,v)
        &\le d(v,p_{i-1}(x))+H(u,p_{i-1}(x)) \\
        &\le \ell(x,v)+h_{i-1}(x)+h_{i-1}(x)+\ell(u,x) \\
        &\stackrel{\eqref{eq:small-hop-h-i-1}}\le d(u,v)+2(i-1)\ell(x,v).
    \end{align*}

    In the remaining case, $h_i(x)>i\ell(x,v)$, and
    \[
        d(x,p_{i-1}(v))
        \stactri\le \ell(x,v)+h_{i-1}(v)
        \stackrel{\eqref{eq:small-hop-h-i-1}}\le \ell(x,v) + (i-1)\ell(x,v) = i\ell(x,v)
        < h_i(x),
    \]
    and therefore $p_{i-1}(v)\in B(x)$.
    Since this case can occur only when $i<k$, the triangle inequality also
    gives $h_i(x)\le h_i(v)+\ell(x,v)$, and hence
    \[
        h_i(v)>(i-1)\ell(x,v)\ge h_{i-1}(v).
    \]
    Thus, $p_{i-1}(v)\in B(v)$, and the query for $(u,v)$ gives
    \begin{align*}
        \hat d(u,v)
        &\stackrel{p_{i-1}(v)\in B(v)}\le H(u,p_{i-1}(v))+d(v,p_{i-1}(v)) \\
        &\stackrel{p_{i-1}(v)\in B(x)}\le \ell(u,x)+d(x,p_{i-1}(v))+h_{i-1}(v) \\
        &\stactri \le d(u,v)+2h_{i-1}(v) \stackrel{\eqref{eq:small-hop-h-i-1}} \le d(u,v) + 2(i-1)\ell(x,v).
    \end{align*}

    Therefore, in either case,
    \[
        \hat d(u,v)
        \le d(u,v)+2(i-1)\ell(x,v) \stackrel{i\leq k}\leq d(u,v)+2(k-1)\ell(x,v)
        \stackrel{\ell(x,v)\leq d(u,v)/2}\le k\,d(u,v).
    \]
    Every estimate considered by the algorithm is the length of a
    $u$--$v$ walk, and hence $d(u,v)\le\hat d(u,v)$, as required.
\end{proof}

%


\section{Distance oracles with nonconstant query time}
\label{sec:dos}

In \Cref{sec:65k}, we presented a $(\frac 65 k+7)$-approximation algorithm for \mPSP that runs in $\Ot(mn^{1/k})$ time. The preprocessing of the algorithm is independent of the set of queried pairs. Therefore, the same construction can be viewed as a distance oracle, where we perform the preprocessing once and evaluate each query using the query procedure of \Cref{alg:65k-alg}.

\Reminder{thm:intro-ado-weighted}

This result is particularly interesting for sparse graphs. When $m=O(n)$, the space becomes $\Ot(n^{1+1/k})$, matching, up to polylogarithmic factors, the size of the classical Thorup--Zwick distance oracle~\cite{DBLP:journals/jacm/ThorupZ05}, while obtaining a substantially smaller stretch at the cost of a larger query time.

\subsection{The unweighted case}

Recall why the table $H$ was needed in \Cref{sec:65k}. In both the Two-Level Growth Lemma and the Long--Short Growth Lemma, the analysis identifies an edge $(x,y)$ of $P(u,v)$. The preprocessing of \Cref{alg:65k-alg} allows the query to connect a pivot associated with $x$ to a vertex in $B(y)$, and therefore to cross the edge $(x,y)$ directly. This is essential in weighted graphs, since the edge $(x,y)$ may have arbitrarily large weight.

In an unweighted graph, every edge has length one. We can therefore choose a vertex of $P(u,v)$ instead of splitting the path across an edge. This changes the inequalities used in the two lemmas only by additive constants. Once the split occurs at a vertex, the required query is of one of the following two forms:
\begin{equation*}
B(u)\times \{p_j(v)\mid 0\le j<k\}
\qquad\text{or}\qquad
\{p_i(u)\mid 0\le i<k\}\times B(v).
\end{equation*}
These are exactly the pairs that we examine during the query.

In the preprocessing phase, after constructing the standard Thorup--Zwick hierarchy, pivots, and bunches, the algorithm updates $H(p_i(x),y)$ and $H(y,p_i(x))$ with the value $h_i(x)+d(x,y)$ for every vertex $x\in V$, every $0\le i<k$, and every $y\in B(x)$, keeping the minimum value for each entry. This takes only $\Ot(n^{1/k})$ time per vertex.

Given a query $(u,v)$, the algorithm returns the minimum of
\[
\min_{w\in B(u), i\in [k]}\{d(u,w)+H(w,p_i(v))+d(p_i(v),v)\}
\]
and
\[
\min_{i\in [k], w\in B(v)} \{d(u,p_i(u))+H(p_i(u),w)+d(w,v) \}.
\]

It remains to explain why the stretch analysis of \Cref{sec:65k} continues to hold. We do not repeat the proofs of the Two-Level Growth Lemma and the Long--Short Growth Lemma. Instead, we describe the only change required in each proof.

\paragraph{The Two-Level Growth Lemma.}
We use the following unweighted version of the lemma.
\begin{lemma}
Let $D=d(u,v)$ and $2\le i<k$. If
\[
h_i(u)-h_{i-2}(v)\ge D+3,
\]
then the query returns an estimate of at most
\[
2h_{i-2}(v)+h_{i-1}(u)+h_i(u)+D.
\]
\end{lemma}

The proof follows the original one, with the additive constant $3$ allowing us to choose a single vertex $x\in P(u,v)$ at which both required strict inequalities hold, as in the proof of the Mixed-Level Growth Lemma. We then have $p_{i-1}(x)\in B(u)$ and $p_{i-2}(v)\in B(x)$, so the required entry $H(p_{i-1}(x),p_{i-2}(v))$ is stored during preprocessing.

\paragraph{The Long--Short Growth Lemma.}
We use the following unweighted version of the lemma.
\begin{lemma}
Let $D=d(u,v)$, $2\le i<k$, and $1\le j<k$. If
\[
\nabla_i(u)+\Delta_j(v)\ge 2D+3,
\]
then the query returns an estimate of at most
\[
2h_{i-1}(u)+2h_{j-1}(v)+3D.
\]
\end{lemma}

The proof follows the original one, with the additive constant $3$ allowing us to choose a single vertex $x\in P(u,v)$ at which both required strict inequalities hold, as in the proof of the Mixed-Level Growth Lemma. The required entry of $H$ is therefore stored during preprocessing using the pivot and bunch associated with this same vertex $x$.

We now apply the numerical analysis of \Cref{sec:65k} with $D'=D+2$ in place of $D$. Since all distances are integral, the conditions $h_i(u)-h_{i-2}(v)>D'$ and $\nabla_i(u)+\Delta_j(v)>2D'$ imply the hypotheses of the two modified lemmas above. Their conclusions remain valid when $D$ is replaced by $D'$, so the same analysis gives
\[
\hat d(u,v)
\le \left(\frac65 k+7\right)D'
= \left(\frac65 k+7\right)d(u,v)+\frac{12}{5}k+14.
\]
Thus, the oracle is a $\left(\frac65 k+7,\frac{12}{5}k+14\right)$-approximation. We therefore obtain \Cref{thm:intro-ado-unweighted}.

\Reminder{thm:intro-ado-unweighted}

\subsection{Hopsets}
\label{sec:hopsets}
Our hopsets consist of the edges used by the walks in our analyses. Each such edge $(a,b)$ gets the length of a walk in $G$ between $a$ and $b$ as its weight, so adding it to $G$ preserves distances. Since our analyses hold for every pair $u,v\in V$, it suffices to count the hops of the walks that they use.

Let $\mathcal{H}_{TZ}$ contain, for every $x\in V$, the edge $(x,w)$ of weight $d(x,w)$ for every $w\in B(x)$, and the edge $(x,p_i(x))$ of weight $h_i(x)$ for every level $i$. Its size is $\Ot(n^{1+1/k})$.

First, we describe our hopset for unweighted graphs.
\paragraph{Four hops in unweighted graphs.}
Set $\mathcal{H}=\mathcal{H}_{TZ}$. Every estimate used in the analysis of \Cref{sec:k-stretch} is the length of a walk $u\to w\to v$ with $w\in B(u)\cap B(v)$, or of a walk $u\to w_u\to x\to w_v\to v$ with $w_u\in B(u)\cap B(x)$ and $w_v\in B(v)\cap B(x)$. Both walks have at most $4$ edges of $\mathcal{H}$ (see \Cref{fig:alg-query}). We obtain:
\Reminder{cor:intro-hopset-unweighted}

For weighted graphs, we have three hopsets.
\paragraph{Three hops in weighted graphs.}
Let $\mathcal{H}$ contain $\mathcal{H}_{TZ}$ together with the entries of the table $H$ of \Cref{alg:65k-alg}. Every estimate in the analysis of \Cref{sec:65k} is then the length of a walk with at most $3$ hops (see \Cref{fig:lb-estimate}), so the stretch guarantee is unchanged. We obtain:
\Reminder{cor:intro-hopset-weighted-3hop}

\paragraph{Five hops in weighted graphs.}
We use the same hopset as in the unweighted case and adapt the analysis of \Cref{sec:k-stretch}. There, the connecting vertex of \Cref{cl:connecting-vertex-exists} is obtained by rounding the ideal split point on $P(u,v)$ to a vertex, and this rounding is the source of the additive $3$ in \eqref{eq:mixed-level-growth-condition}.
In a weighted graph, we do not round. If $\nabla_i(u)+\nabla_j(v)>2D$ and $\nabla_i(u),\nabla_j(v)\le 2D$,\footnote{Otherwise, the first claim in the proof of \Cref{lem:mixed-level-growth} gives a walk of the required length with $2$ hops.} we take the edge $(x,y)$ of $P(u,v)$ that contains the ideal split point, as in \Cref{cl:long-short-split-edge}, so that $\nabla_i(u)>2d(u,x)$ and $\nabla_j(v)>2d(y,v)$.
Applying \Cref{cl:pivot-connection} at $x$, and symmetrically at $y$, gives $w_u\in B(u)\cap B(x)$ and $w_v\in B(v)\cap B(y)$, and hence the walk
\[
u\to w_u\to x\to y\to w_v\to v
\]
of length at most $2h_{i-2}(u)+2h_{j-2}(v)+3D$. This walk has $5$ hops: four hopset edges and the edge $(x,y)$ of $G$.
Since the condition is now $\nabla_i(u)+\nabla_j(v)>2D$ instead of \eqref{eq:mixed-level-growth-condition}, the numerical argument of \Cref{sec:k-stretch} (the proof of \Cref{lem:k-optimization} with $L=2D$, using the condition $x_i+y_j>2D$ in place of $x_i+y_j\ge2D+3$, which turns the bounds $\kappa(D+1)$ and $(\kappa-1)(D+1)$ into $\kappa D$ and $(\kappa-1)D$) gives stretch $k$ without the additive $k-1$.
\Reminder{cor:intro-hopset-weighted}

\paragraph{Four hops in weighted graphs.}
The edge $(x,y)$ and the hop from $y$ to $w_v$ can be merged into a single hopset edge. Let $\mathcal{H}$ contain $\mathcal{H}_{TZ}$ and, for every oriented edge $(x,y)$ of $G$ and every $z\in B(y)$, the edge $(x,z)$ of weight $\wt(x,y)+d(y,z)$. There are $\Ot(mn^{1/k})$ such edges. The walk above becomes $u\to w_u\to x\to w_v\to v$, which has $4$ hops and the same length, and we obtain:
\Reminder{cor:intro-hopset-weighted-4hop}

\section{Generalizing to $(t,s)$-PSP}
\label{sec:ts-psp}
In this section, we extend the results of the previous sections to the more general $(t,s)$-PSP problem, which has $s$ sources for the $t$ pairs of query vertices, i.e., the input is a set $I$ of $t$ pairs together with a set $A\subseteq V$ of size $s$ such that $A\cap \{u,v\}\neq \emptyset$ for every pair $\{u,v\}\in I$.

Notice that if $s=n$, then this problem is exactly \tPSP. In this section, we show that if $s\ll n$, then improved algorithms are obtained.
In fact, essentially every prescribed-pairs problem studied in this paper is a special case of $(t,s)$-PSP: \nPSP is $(n,n)$-PSP, \mPSP is $(m,n)$-PSP, and \tPSP is $(t,\min\{t,n\})$-PSP, since one can always take $A$ to contain one endpoint of every pair. The $ST$-shortest paths problem with $I\subseteq S\times T$ and $|S|\le |T|$ is $(|I|,|S|)$-PSP, and MSSP with a set $S$ of sources is $(|S|\cdot n,|S|)$-PSP.

Our extension to $(t,s)$-PSP relies on the following idea, previously used in the parameterized approximate distance oracles of~\cite{kadria2025fasteralgorithms2k1stretchdistance, RodittyTZ05}.
Instead of a hierarchy $A_0=V\supseteq A_1\supseteq \dots \supseteq A_k=\emptyset$, each sampled with probability $n^{-1/k}$, we let $A_0=A$ be the set of sources of the queries.
The rest of the hierarchy is then obtained by sampling every vertex of $A_{i-1}$ with probability $s^{-1/k}$.

The set $A$ need not be given as part of the input, and a near-optimal one can be computed in linear time.
A smallest set that hits every pair of $I$ is exactly a minimum vertex cover of the graph $(V,I)$. Computing it is NP-hard, but a $2$-approximation is enough for our purposes.
We scan the pairs of $I$ in an arbitrary order and, whenever we encounter a pair $\{u,v\}$ that is not yet hit by $A$, we add both $u$ and $v$ to $A$.
The pairs added this way are pairwise disjoint, and every hitting set contains an endpoint of each of them. Hence, $|A|\le 2s$, and the whole computation takes $O(t)$ time.
As $s$ enters our bounds only through $s^{1/k}$ and $s^{2/k}$, this loses a factor of at most $2^{2/k}=1+O(1/k)$ in the running time.

We then use the construction of~\cite{DBLP:journals/jacm/ThorupZ05} to compute the bunches with respect to the new hierarchy in $\Ot(ms^{1/k})$ time, and continue to run the algorithms of the previous sections with respect to the new hierarchy and bunches. For the new hierarchy, we have $|B(w)|=\Ot(s^{1/k})$ for every $w\in V$.

Notice that for an input pair $\{u,v\}\in I$, we may assume without loss of generality that $u\in A_0$, and hence $h_0(u)=0$ and $h_0(v)\le d(v,u)+h_0(u)=d(u,v)$.
The only difference from the previous sections is therefore that the hierarchy no longer starts at $h_0(u)=h_0(v)=0$.
The analysis tools of the previous sections, for both weighted and unweighted graphs, are carefully designed so that their assumptions involve only the increments $\Delta_i(\cdot)$ and $\nabla_i(\cdot)$, which are unchanged by subtracting the initial heights, and, in the weighted case, differences between heights at the two endpoints, while their conclusions bound $\hat d(u,v)$ by $2h_i(u)+2h_j(v)$ plus an additive term.
Therefore, running the same algorithms with the new hierarchy adds at most $2h_0(u)+2h_0(v)\le 2d(u,v)$ to the estimate in the unweighted case. In the weighted case, the differences between the two endpoints require the additional adjustment below, which adds at most $4h_0(v)\le4d(u,v)$ to the estimate.
We therefore obtain the following two theorems for the general $(t,s)$-PSP problem:
\Reminder{thm:intro-tsPSP-unweighted}
\begin{proof}
    As discussed above, since $h_0(u)=0$ and $h_0(v)\le d(u,v)$, the stretch analysis of \Cref{sec:k-stretch}, which relies only on the increments $\nabla_i(u)$ and $\nabla_j(v)$ through the Mixed-Level Growth Lemma (\Cref{lem:mixed-level-growth}) and the numerical optimization of \Cref{lem:k-optimization}, follows with an additive $2d(u,v)$ in the estimate.
    In more detail, the proof of \Cref{lem:mixed-level-growth} uses only \Cref{lem:tz-bunch} and the triangle inequality, so it holds for any hierarchy, and its condition involves only $\nabla_i(u)$ and $\nabla_j(v)$.
    In \Cref{sec:k-stretch}, the numerical argument writes $h_{2i-2}(u)=\sum_{r<i}x_r$ and $h_{2j-2}(v)=\sum_{r<j}y_r$, which uses $h_0(u)=h_0(v)=0$.
    In general, $h_{2i-2}(u)=h_0(u)+\sum_{r<i}x_r$ and $h_{2j-2}(v)=h_0(v)+\sum_{r<j}y_r$.
    Hence, both bounds used there, the bound of \Cref{lem:mixed-level-growth} and the last-level bound $2\min\{h_{k-1}(u),h_{k-1}(v)\}+d(u,v)$, exceed the expressions bounded in \Cref{lem:k-optimization} by at most $2h_0(u)+2h_0(v)\le 2d(u,v)$.
    This gives the $(k+2,k-1)$-approximation.

    It remains to bound the running time. Constructing the bunches takes $\Ot(ms^{1/k})$ time.
    Constructing $H$ takes $\sum_{w\in V}|B(w)|^2=\Ot(n\cdot s^{2/k})$ time. Querying a pair $\{u,v\}\in I$ takes $|B(u)|\cdot |B(v)|=\Ot(s^{2/k})$ time. Therefore, querying all $t$ pairs takes $\Ot(ts^{2/k})$ time.
    Thus, the total running time is $\Ot(ms^{1/k}+(n+t)s^{2/k})$, as required.
\end{proof}

\Reminder{thm:intro-tsPSP-weighted}
\begin{proof}
    As discussed above, since $h_0(u)=0$ and $h_0(v)\le d(u,v)$, the stretch analysis of \Cref{sec:65k} follows with an additive $4d(u,v)$ in the estimate.
    In more detail, we apply the numerical optimization of \Cref{lem:six-fifths-optimization} to the heights $\bar h_i(w)=\max\{h_i(w)-h_0(v),0\}$ for $w\in\{u,v\}$, so that $\bar h_0(u)=\bar h_0(v)=0$.
    This replacement cannot increase the increments $\Delta_i(\cdot)$ and $\nabla_i(\cdot)$, or any positive difference between heights at the two endpoints. Therefore, whenever the condition of the Two-Level Growth Lemma (\Cref{lem:two-level-growth}) or the Long--Short Growth Lemma (\Cref{lem:long-short}) holds for the modified heights, it also holds for the original heights.
    Since $h_i(w)\le\bar h_i(w)+h_0(v)$, all the bounds used in \Cref{sec:65k} exceed their expressions in the modified heights by at most $4h_0(v)\le4d(u,v)$.
    This gives the $(\frac65 k+11)$-approximation.

    It remains to bound the running time. Constructing the bunches takes $\Ot(ms^{1/k})$ time.
    Constructing $H$ takes $\sum_{(x,y)\in E}k\cdot|B(y)|=\Ot(m\cdot s^{1/k})$ time, where the sum is over the oriented edges of $G$.
    Querying a pair $\{u,v\}\in I$ takes $k\cdot|B(v)|+|B(u)|\cdot k=\Ot(s^{1/k})$ time. Therefore, querying all $t$ pairs takes $\Ot(ts^{1/k})$ time. Thus, the total running time is $\Ot((m+t)s^{1/k})$, as required.
\end{proof}

Instantiating \Cref{thm:intro-tsPSP-unweighted,thm:intro-tsPSP-weighted} with the special cases listed above recovers all of our prescribed-pairs bounds.
Taking $s=n$ together with $t=n$ and $t=m$, respectively, recovers the running times of \Cref{thm:intro-unweighted-npsp,thm:intro-weighted}, where the additive terms $2d(u,v)$ and $4d(u,v)$, respectively, are not paid since $A=V$ implies $h_0(u)=h_0(v)=0$.
Taking $s=O(t)$ gives the improved \tPSP bounds of \Cref{cor:intro-tPSP-weighted,cor:intro-tPSP-unweighted} for $t\ll n$, taking $s=|S|$ and $t=|I|$ gives \Cref{cor:intro-STSP-weighted,cor:intro-STSP-unweighted} for $ST$-SP, and taking $s=|S|$ and $t=|S|\cdot n$ gives MSSP in $\Ot((m+|S|n)|S|^{1/k})$ time for weighted graphs and in $\Ot(m|S|^{1/k}+n|S|^{1+2/k})$ time for unweighted graphs.

\bibliography{bibliography}
\clearpage
\appendix

\section{Numerical optimization lemmas}\label{app:numerical-lemma}

\subsection{Solving the optimization problem for stretch $k$}
In this section, we prove the following lemma, used in the proof of \Cref{thm:intro-unweighted-npsp}.

\Reminder{lem:k-optimization}

We first prove a normalized version of the lemma, in which all increments lie in $[0,1]$, and then derive \Cref{lem:k-optimization} from it by capping and scaling.
\begin{claim}\label{cl:k-unit-gap-normalized}
Let $\kappa\ge0$ and $\beta\ge0$, and let
$x_1,\ldots,x_\kappa,y_1,\ldots,y_\kappa\in[0,1]$.
Write
\(
    X_i=\sum_{r=1}^i x_r,
    Y_j=\sum_{r=1}^j y_r,
\)
and set $X_0=Y_0=0$.
Then either
\(
    X_\kappa+Y_\kappa\le\kappa
\),
or
\(
    \min\{X_\kappa,Y_\kappa\}\le\kappa/2-\beta,
\)
or there are $i,j\ge1$ with $i+j\le\kappa+1$ such that
\[
    x_i+y_j>1
    \qquad\text{and}\qquad
    X_{i-1}+Y_{j-1}\le\frac{\kappa-1}{2}+\beta.
\]
\end{claim}

\begin{proof}
We call a pair of indices $(i,j)\in\{1,\ldots,\kappa\}^2$ a \emph{crossing}
if $x_i+y_j>1$.
We may assume that both sequences are nondecreasing: replace each
entry by the maximum of all entries up to its position.
Write $\bar x_i,\bar y_j$ for the modified entries and
$\bar X_i,\bar Y_j$ for their prefix sums. Complete sums can only
increase, so the first two alternatives for the modified sequences imply
the same alternatives for the original sequences. For a crossing $(i,j)$ in the modified
sequences, choose $a\le i$ and $b\le j$ attaining the running maxima,
so $x_a=\bar x_i$ and $y_b=\bar y_j$. Then
\[
    x_a+y_b>1,\qquad a+b\le i+j,\qquad
    X_{a-1}+Y_{b-1}\le\bar X_{i-1}+\bar Y_{j-1}.
\]
Thus the third alternative also transfers to the original sequences.
We henceforth work with the modified sequences, retaining the original notation.

Suppose the conclusion fails. In particular,
$X_\kappa+Y_\kappa>\kappa$ and $X_\kappa,Y_\kappa>\kappa/2-\beta$.
If every pair with $i+j\le\kappa+1$ satisfied $x_i+y_j\le1$,
then summing $x_i+y_{\kappa+1-i}\le1$ over $1\le i\le\kappa$
would give
\[
    X_\kappa+Y_\kappa\le\kappa,
\]
a contradiction. Hence, there is a crossing with $i+j\le\kappa+1$.

Let $r$ be the minimum value of $i+j$ over all crossings.
Thus $r\le\kappa+1$. If $r=2$, then $(1,1)$ is a crossing
and $X_0+Y_0=0$, a contradiction. We may therefore assume $r\ge3$.
Minimality gives
\[
    x_a+y_{r-1-a}\le1
    \qquad\text{for every }1\le a\le r-2.
\]
Summing, we obtain
\[
    X_{r-2}+Y_{r-2}\le r-2.
\]
By symmetry, assume $X_{r-2}\le Y_{r-2}$, so
$X_{r-2}\le(r-2)/2$.

\paragraph{If there is a crossing with $i+j=r$ and $i\ge j$.}
Since both sequences are nondecreasing, their prefix averages
are nondecreasing. Therefore,
\begin{align*}
    X_{i-1}+Y_{j-1}
    &\le
    \frac{i-1}{r-2}X_{r-2}
    +\frac{j-1}{r-2}Y_{r-2}\\
    &=
    \frac{j-1}{r-2}(X_{r-2}+Y_{r-2})
    +\frac{i-j}{r-2}X_{r-2}\\
    &\le j-1+\frac{i-j}{2}\\
    &=\frac{r-2}{2}
    \le\frac{\kappa-1}{2}+\beta,
\end{align*}
a contradiction.

\paragraph{Otherwise.}
Every crossing with $i+j=r$ satisfies $i<j$.
Choose such a crossing with $i$ as large as possible.
For every $1\le a<j$, we have $r-a>i$, so this choice implies
$x_{r-a}+y_a\le1$. Summing these inequalities gives
\[
    X_{r-1}-X_i+Y_{j-1}\le j-1.
\]
Using the nondecreasing prefix averages of the $x$-sequence,
we obtain
\begin{equation}\label{eq:k-opt-prefix-bound}
\begin{aligned}
    X_{i-1}+Y_{j-1}
    &\le j-1-X_{r-1}+X_i+X_{i-1}\\
    &\le j-1-\frac{j-i}{r-1}X_{r-1}.
\end{aligned}
\end{equation}
Since each $x_a$ is at most $1$ and $X_\kappa>\kappa/2-\beta$,
\[
    X_{r-1}
    \ge X_\kappa-(\kappa-r+1)
    >r-1-\frac{\kappa}{2}-\beta.
\]
Since $j>i$, we can substitute this lower bound
into \eqref{eq:k-opt-prefix-bound}, obtaining
\begin{align*}
    X_{i-1}+Y_{j-1}
    &<j-1-\frac{j-i}{r-1}
        \left(r-1-\frac{\kappa}{2}-\beta\right)\\
    &=i-1+\frac{\kappa(j-i)}{2(r-1)}+\frac{\beta(j-i)}{r-1}\\
    &=\frac{\kappa-1}{2}
      -\frac{(2i-1)(\kappa-r+1)}{2(r-1)}+\frac{\beta(j-i)}{r-1}\\
    &\le\frac{\kappa-1}{2}+\beta.
\end{align*}
The last inequality uses $r\le\kappa+1$ and $j-i<r-1$.
This is again a contradiction.
\end{proof}

\begin{proof}[Proof of~\Cref{lem:k-optimization}]
Let $L=2D+2$. The entries may exceed $L$, so we first cap them.
For every $1\le r\le\kappa$, let $x'_r=\min\{x_r,L\}$ and $y'_r=\min\{y_r,L\}$.
We apply \Cref{cl:k-unit-gap-normalized} with $\beta=c/L$ to the sequences $x'_r/L$ and $y'_r/L$, whose entries lie in $[0,1]$.
Since the entries are integers, a crossing $x'_i+y'_j>L$ of the normalized sequences implies $x'_i+y'_j\ge2D+3$.
For a pair $(i,j)$ with $x_i+y_j\ge2D+3$, we call $\sum_{r<i}x_r+\sum_{r<j}y_r+D$ its \emph{value}.
Multiplying the resulting bounds by $L$, we obtain one of the following two cases.

\paragraph{The first two alternatives.}
We have $\sum_{r=1}^{\kappa}x'_r+\sum_{r=1}^{\kappa}y'_r\le2\kappa(D+1)$ or $\min\{\sum_{r=1}^{\kappa}x'_r,\sum_{r=1}^{\kappa}y'_r\}\le\kappa(D+1)-c$. In the first case, if $\sum_{r=1}^{\kappa}x'_r>\kappa(D+1)+c$, then $\sum_{r=1}^{\kappa}y'_r<\kappa(D+1)-c$. Hence, in both cases,
\[
    \sum_{r=1}^{\kappa}x'_r\le\kappa(D+1)+c
    \qquad\text{or}\qquad
    \sum_{r=1}^{\kappa}y'_r+2c\le\kappa(D+1)+c.
\]
Suppose first that $\sum_{r=1}^{\kappa}x'_r\le\kappa(D+1)+c$.
If no entry of the $x$-sequence was capped, then $\sum_{r=1}^{\kappa}x_r\le\kappa D+\kappa+c$.
Otherwise, let $a$ be the smallest index with $x_a>L$. Then $x_a\ge2D+3$, so the pair $(a,1)$ satisfies $x_a+y_1\ge2D+3$.
Since $x'_a=L$ and no entry before $a$ was capped,
\[
    \sum_{r<a}x_r=\sum_{r<a}x'_r\le\kappa(D+1)+c-L,
\]
and the pair $(a,1)$ gives a value of at most $\kappa(D+1)+c-L+D<\kappa D+\kappa+c$.
The case $\sum_{r=1}^{\kappa}y'_r+2c\le\kappa(D+1)+c$ is analogous: if no entry of the $y$-sequence was capped, then $\sum_{r=1}^{\kappa}y_r+2c\le\kappa D+\kappa+c$, and otherwise the smallest capped index $a$ gives the pair $(1,a)$, whose value is at most $\kappa(D+1)-c-L+D<\kappa D+\kappa+c$.

\paragraph{The third alternative.}
There are $i,j$ with $x'_i+y'_j\ge2D+3$ and
\[
    \sum_{r<i}x'_r+\sum_{r<j}y'_r\le(\kappa-1)(D+1)+c.
\]
Since $x_i\ge x'_i$ and $y_j\ge y'_j$, the pair $(i,j)$ also satisfies $x_i+y_j\ge2D+3$.
If no entry $x_r$ with $r<i$ and no entry $y_r$ with $r<j$ was capped, the prefix sums are unchanged, and the pair $(i,j)$ gives a value of at most
\[
    \sum_{r<i}x_r+\sum_{r<j}y_r+D
    \le(\kappa-1)(D+1)+c+D
    =\kappa D+\kappa+c-1.
\]
Otherwise, some $x_r$ with $r<i$ or some $y_r$ with $r<j$ was capped. Since the bound on the prefix sums is symmetric, we may assume the former. Let $a$ be the smallest such index. Then $x_a\ge2D+3$, so the pair $(a,1)$ satisfies $x_a+y_1\ge2D+3$, and
\[
    \sum_{r<a}x_r=\sum_{r<a}x'_r\le\sum_{r<i}x'_r\le(\kappa-1)(D+1)+c.
\]
Hence the pair $(a,1)$ gives a value of at most $\kappa D+\kappa+c-1$.

In every case, the required minimum is at most
$\kappa D+\kappa+c$, as claimed.
\end{proof}

\pagebreak

\subsection{The $\frac{6}{5}k$ case}\label{app:six-fifths-optimization}

\Reminder{lem:six-fifths-optimization}

\begin{proof}
Let $\sol$ denote the minimum in the statement. We prove the
bound $\sol\le \frac35 k+2$. The case $\sol\le3$ is
immediate. Also, if $k=2$, then $\sol\le X_1\le2$.
Thus assume that $k\ge3$ and $\sol>3$.

We will choose a cutoff index $q$ and prove
\begin{equation*}
q\ge\sol-4,
\qquad
8(\sol-1)\le9(k-1)-7q.
\end{equation*}
The first inequality follows from the two-level terms in the
minimum. To obtain the second, we encode the tail of one sequence
and the prefix of the other by intervals, and use the long--short
terms to bound the lengths of overlapping intervals.

Replace each $X_i$ by $\min\{X_i,\sol\}$ and each $Y_i$ by
$\min\{Y_i,\sol\}$, and recompute the increments. We retain the
same notation, keeping $\sol$ fixed as the original minimum.
Thus $X_{k-1}=Y_{k-1}=\sol$.

\begin{claim}[Local restrictions]
\label{cl:six-fifths-local}
For $2\le i\le k-1$ and $1\le j\le k-1$,
\begin{equation}
\label{eq:six-fifths-local-condition}
X_{i-1}+Y_{j-1}<\sol
\quad\Longrightarrow\quad
x_{i-1}+x_i+y_j\le2,
\end{equation}
and the same implication holds after exchanging $X,x$ and $Y,y$.
Moreover, for $2\le i\le k-1$,
\begin{equation}
\label{eq:six-fifths-prefix-growth}
X_i+Y_i<\sol
\quad\Longrightarrow\quad
X_i+Y_i\le X_{i-2}+Y_{i-2}+2.
\end{equation}
Every increment and every pair of consecutive increments is at
most $2$, and $X_1+Y_1\le2$.
\end{claim}

\begin{proof}
If the hypothesis of
\eqref{eq:six-fifths-local-condition} holds after truncation,
both prefixes are below $\sol$ and hence unchanged. A violation
of its conclusion would also have held before truncation,
since increments only decrease. It would therefore make
$X_{i-1}+Y_{j-1}$ a feasible objective below $\sol$,
contradicting the definition of the minimum.

Similarly, if $X_i+Y_i<\sol$, these prefixes are unchanged.
Both two-level objectives at index $i$ are at most $X_i+Y_i$,
so neither of their defining strict conditions can hold.
Hence
\begin{equation*}
X_i\le Y_{i-2}+1,
\qquad
Y_i\le X_{i-2}+1.
\end{equation*}
Adding these inequalities proves
\eqref{eq:six-fifths-prefix-growth}.

Every positive capped increment $x_i$ starts at
$X_{i-1}<\sol$. For $i\ge2$, applying
\eqref{eq:six-fifths-local-condition} with $j=1$ gives $x_i\le2$.
The first increment is bounded by assumption, and the same
argument applies to $Y$.

For $1\le i\le k-2$, if $X_i<\sol$, apply
\eqref{eq:six-fifths-local-condition} at $(i+1,1)$ to obtain
$x_i+x_{i+1}\le2$. If $X_i=\sol$, then $x_{i+1}=0$, so the
same bound follows from $x_i\le2$. The argument for $Y$ is identical.

Finally, $X_1=x_1\le2<\sol$, so
\eqref{eq:six-fifths-local-condition} at $(2,1)$ gives
$x_1+x_2+y_1\le2$, and therefore $X_1+Y_1\le2$.
\end{proof}

We now locate the cutoff. Let $q$ be the largest index such that
$X_q+Y_q<\sol-1$, and write
\begin{equation*}
S=X_q+Y_q.
\end{equation*}

\begin{claim}[The cutoff index]
\label{cl:six-fifths-cutoff}
\begin{equation*}
1\le q\le k-2,
\qquad
q\ge\sol-4.
\end{equation*}
\end{claim}

\begin{proof}
By \Cref{cl:six-fifths-local},
$X_1+Y_1\le2<\sol-1$, whereas
$X_{k-1}+Y_{k-1}=2\sol$. Thus $1\le q\le k-2$.

For every $2\le i\le q$, we have $X_i+Y_i\le S<\sol$.
Hence \eqref{eq:six-fifths-prefix-growth} shows that every two
levels add at most $2$ to the combined prefix sum.
Starting from $X_0+Y_0=0$ on the even indices and
$X_1+Y_1\le2$ on the odd indices gives
\begin{equation*}
S\le q+1.
\end{equation*}

Since $S<\sol$, applying
\eqref{eq:six-fifths-local-condition} at $i=j=q+1$ gives
$x_q+x_{q+1}+y_{q+1}\le2$. In particular,
$x_{q+1}+y_{q+1}\le2$.
The maximality of $q$ therefore implies
\begin{equation*}
\sol-1
\le X_{q+1}+Y_{q+1}
\le S+2
\le q+3,
\end{equation*}
which yields $q\ge\sol-4$.
\end{proof}

Next, we compare the remaining growth of $X$ with the accumulated
growth of $Y$. By symmetry, assume that $X_q\le Y_q$.
Then
\begin{equation*}
(\sol-X_q)+Y_q\ge\sol.
\end{equation*}
We encode these two quantities by intervals, with a loss of at most
$1$ in their total length.

The long--short restriction involves two consecutive increments
of $X$ and one increment of $Y$. Accordingly, define
\begin{equation*}
\begin{aligned}
I_i
&=
\left[
    \frac{X_{i-1}+X_i}{2},
    \frac{X_i+X_{i+1}}{2}
\right]
&& (q\le i\le k-2),\\
J_j
&=
[\sol-1-Y_j,\ \sol-1-Y_{j-1}]
&& (1\le j\le q).
\end{aligned}
\end{equation*}
Thus $|I_i|=(x_i+x_{i+1})/2$ and $|J_j|=y_j$.
The midpoint endpoints make the $I_i$ consecutive, while the
reflection of the $J_j$ will make overlap imply the prefix
condition in \eqref{eq:six-fifths-local-condition}.

\begin{claim}[The interval representation]
\label{cl:six-fifths-intervals}
Each family has disjoint interiors, and every $J_j$ lies inside
$\bigcup_i I_i$. Moreover,
\begin{equation}
\label{eq:six-fifths-short-length}
\sum_i |I_i|+\sum_j |J_j|\ge\sol-1.
\end{equation}
Every $I_i$ has length at most $1$, and positive overlap of
$I_i$ and $J_j$ implies
\begin{equation}
\label{eq:six-fifths-overlap}
2|I_i|+|J_j|\le2.
\end{equation}
\end{claim}

\begin{proof}
The intervals within each family are consecutive. Their unions are
\begin{equation*}
\bigcup_i I_i
=
\left[
    X_q-\frac{x_q}{2},
    \sol-\frac{x_{k-1}}{2}
\right],
\qquad
\bigcup_j J_j
=
[\sol-1-Y_q,\ \sol-1].
\end{equation*}
Since $S<\sol-1$, the left endpoint of the second union is greater
than $X_q$. Since $x_{k-1}\le2$, its right endpoint is at most the
right endpoint of the first union. This proves the containment.
The threshold $\sol-1$ in the definition of $q$ was chosen to
ensure precisely this property.

Summing the lengths gives
\begin{equation*}
\sum_i |I_i|+\sum_j |J_j|
=
\sol-X_q+Y_q+\frac{x_q-x_{k-1}}{2}
\ge\sol-1,
\end{equation*}
using $Y_q\ge X_q$, $x_q\ge0$, and $x_{k-1}\le2$.

\Cref{cl:six-fifths-local} gives
$|I_i|=(x_i+x_{i+1})/2\le1$.
If $I_i$ and $J_j$ overlap in positive length, the left endpoint
of $I_i$ is below the right endpoint of $J_j$, so
\begin{equation*}
\frac{X_{i-1}+X_i}{2}<\sol-1-Y_{j-1}.
\end{equation*}
Since $x_i/2\le1$, this implies
\begin{equation*}
X_i+Y_{j-1}<\sol-1+\frac{x_i}{2}\le\sol.
\end{equation*}
Applying \eqref{eq:six-fifths-local-condition} at $(i+1,j)$ yields
\begin{equation*}
2|I_i|+|J_j|
=
x_i+x_{i+1}+y_j
\le2,
\end{equation*}
as required.
\end{proof}

The remaining task is to turn the total length in
\Cref{cl:six-fifths-intervals} into a bound on the number
of intervals. We count the lengths of the two families separately:
a point in their overlap contributes twice to the total length.

\begin{claim}[Counting the intervals]
\label{cl:six-fifths-counting}
\begin{equation*}
8\left(\sum_i |I_i|+\sum_j |J_j|\right)
\le9(k-1-q)+2q.
\end{equation*}
\end{claim}

\begin{proof}
We assign weights that provide at least $8$ per unit of length
in each family. Thus a point covered by both families must receive
density at least $16$.

Distribute weight $9$ uniformly over each nonempty $I_i$, and
weight $2$ uniformly over each nonempty $J_j$.
At a point covered only by an $I_i$, the density is
$9/|I_i|\ge9\ge8$.
On a positive overlap, \eqref{eq:six-fifths-overlap} and
Cauchy--Schwarz give
\begin{align*}
\frac9{|I_i|}+\frac2{|J_j|}
&=
2\left(
    \frac{3^2}{2|I_i|}
    +\frac{1^2}{|J_j|}
\right)\\
&\ge
\frac{2(3+1)^2}{2|I_i|+|J_j|}
\ge16.
\end{align*}

Apart from endpoints, each point belongs to at most one interval
from each family. There are no points covered only by a $J_j$,
because every $J_j$ lies inside $\bigcup_i I_i$.
Hence the total weight is at least
\begin{equation*}
8\left(\sum_i |I_i|+\sum_j |J_j|\right).
\end{equation*}
There are $k-1-q$ intervals $I_i$ and $q$ intervals $J_j$, so the
assigned weight is at most $9(k-1-q)+2q$. This proves the claim.
\end{proof}

\Cref{cl:six-fifths-intervals} provides a total length of at least
$\sol-1$, and \Cref{cl:six-fifths-counting} bounds the weight
needed to cover that length. The resulting budget is
$9(k-1-q)+2q=9(k-1)-7q$, so the lower bound on $q$ from
\Cref{cl:six-fifths-cutoff} now gives
\begin{equation*}
8(\sol-1)
\le
9(k-1)-7q
\le
9(k-1)-7(\sol-4).
\end{equation*}
Therefore $15\sol\le9k+27$, and hence
\begin{equation*}
\sol\le\frac35 k+\frac95<\frac35 k+2.
\end{equation*}
\end{proof}

\end{document}